\documentclass[showpacs,floatfix,a4paper]{revtex4}
\usepackage{epsfig,color}
\usepackage{amsmath}
\usepackage{amssymb}
\usepackage{graphicx}
\usepackage{here}
\newcommand{\feyn}[1]{{#1}\!\!\!{\slash}}
\newcommand{\pa}{\ensuremath{P}}
\newcommand{\pb}{\ensuremath{q}}
\newcommand{\pone}{\ensuremath{P'}}
\newcommand{\pthree}{\ensuremath{q'}}
\newcommand{\mQ}{\ensuremath{M}}
\newcommand{\mQsq}{\ensuremath{M^2}}
\newcommand{\kt}{\ensuremath{k_t}}
\newcommand{\lt}{\ensuremath{\ell_t}}
\newcommand{\vkt}{\ensuremath{\vec k_t}}
\newcommand{\vlt}{\ensuremath{\vec\ell_t}}
\newcommand{\alphas}{\ensuremath{\alpha_s}}
\newcommand{\mD}{\ensuremath{\mu}}

\begin{document}

\title{Gluon radiation by heavy quarks at intermediate energies}

\author{Joerg Aichelin, Pol Bernard Gossiaux, Thierry Gousset}

\affiliation{SUBATECH, CNRS/IN2P3, Universit\'e de Nantes, Ecole des
  Mines de Nantes\\ 4 rue Alfred Kastler, 44307 Nantes cedex 3,
  France}

\date{\today}

\begin{abstract}
Employing scalar QCD we study the gluon emission of heavy quarks
created by the interaction with light quarks considered as
dynamical scattering centers. We develop approximation
formulas for the high energy limit and study when the full calculation
reaches this high energy limit. For zero quark masses and in the high
energy limit our model reproduces the Gunion-Bertsch results. We
justify why scalar QCD represents a good approximation to the full QCD
approach for the energy loss of heavy quarks. In the regime of
accessible phenomenology we observe that the emission at small
transverse momentum (dead cone effect) is less suppressed than
originally suggested. We also investigate the influence of a finite
gluon mass on the discussed results.
\end{abstract}

\maketitle

\section{Introduction}

Lattice Gauge Theory predicts that at high temperatures/densities a new state of matter is formed,
a plasma of quarks and gluons (QGP) \cite{Borsanyi:2010cj}. There is strong circumstantial evidence 
that in ultrarelativistic heavy-ion collisions such a plasma  is created for a short amount of time.
It quickly expands and hadronizes. It is the main objective of the present experiments
at the ultrarelativistic heavy ion colliders to study the properties of the QGP. 
The experiments of the last ten years at RHIC as well as the first runs at LHC 
have revealed that the hadron multiplicities are compatible with
the assumption that hadrons are produced in statistical equilibrium at
a temperature compatible with the predictions of Lattice Gauge
calculations for the chiral/confinement phase transition \cite{Borsanyi:2010cj}.
Therefore the hadrons which are formed from plasma constituents are only
of very limited use for the understanding of properties of the QGP.

For the study of the properties of the QGP during its expansion one
has to rely on probes which do not come to an equilibrium with the
plasma constituents. High-momentum heavy hadrons, those which contain
a charm or a bottom quark, are such a probe. Due to the high energy
required for their production heavy quarks are created in hard collisions 
during the initial phase of the reaction and do not annihilate in later 
phases~\cite{Andronic:2007}. The number of these
collisions can be determined from the collision geometry and the
initial momentum distribution of the heavy quarks can be calculated
from perturbative QCD (pQCD)~\cite{Vogt:2003,Cacciari:2005,Cacciari:2012}.
During the expansion of the plasma the
heavy quarks interact with the plasma constituents, light quarks and
gluons, but their initial momentum distribution is so different from
that of the plasma particles that they do not come to thermal
equilibrium~\cite{Phenix:2011,ALICE:2012}. Therefore, their final momentum distribution at
hadronization contains the desired information on the properties of
the plasma during its evolution and this information is transferred to
the heavy hadrons whose kinematics is largely determined by that of
the entrained heavy quark.

The interpretation of the experimental (open) heavy flavor results is in
reality a double challenge: One has to understand the elementary
interaction of the heavy quarks with the plasma constituents but also
the expansion of the plasma itself. For the same elementary
interaction different expansion scenarios yield different results of
the observables~\cite{Gossiaux:2011ea}.

Heavy quarks interact with the plasma constituents either by elastic
collisions \cite{Bjorken,Braaten:1991jj,Peshier:2006hi,Peigne:2008nd,Gossiaux:2008jv}
or by inelastic radiative collisions
\cite{Gyulassy94,Wang95,Baier95,Baier97,Zakharov,GLV,Dokshitzer:2001zm,AMY,Zakharov:2004,ASW,Zhang:2004,Djordjevic:2004}
or both \cite{Wicks:2007am}. Whereas
radiative collisions dominate the energy loss of light quarks, for the
heavy quarks the relative importance of the elastic and of the
radiative energy loss is debated. Detailed calculations for an
expanding plasma are not available yet and the approximate
calculations, using a static plasma of a given length, indicate that
both are of the same order of magnitude ~\cite{Wicks:2007am}. Another
complication for the judgement of the importance of the radiative
energy loss is the Landau Pomeranchuck Migdal (LPM) effect, which
states that radiative collisions are not independent but that a second
gluon can only be emitted after the first one is formed.

For energetic light quarks the LPM effect in an infinite medium with a
constant temperature and with static scattering centers has been
evaluated independently by Zakharov~\cite{Zakharov} and by
Baier, Dokshitzer, Mueller, Peign\'e and Schiff~\cite{Baier95,Baier97}.
Later it has been found that both approaches are identical~\cite{Baier:2000mf} 
and the approach has been extended to an expanding
medium by applying time-dependent transport coefficients
\cite{Baier:1998yf} or time-dependent parton densities~\cite{Zakharov:98}. 
More recently Arnold, Moore and
Yaffe~\cite{Arnold:2002ja}, using diagrammatic methods, extended these
calculations to dynamical gauge fields. The influence of the LPM
effect for heavy quarks in a static medium is, however, presently
still under debate and the calculation of how it shows up in an
expanding medium whose temperature is rapidly changing is a
theoretical challenge which has not been met yet.

A while ago we have advanced a pQCD-inspired calculation for the elastic
collisions of heavy quarks with the QGP constituents which employs a
running coupling constant and an infrared regulator which reproduces
the energy loss of the heavy quarks in the hard thermal loop
approach~\cite{Gossiaux:2008jv,Gossiaux:2009mk}. Embedding these cross
sections in the hydrodynamical description of the expanding plasma of
Heinz and Kolb~\cite{Kolb:2003dz} we could show that the collisional
energy loss underpredicts the measured energy loss of heavy mesons at
large momenta as well as their elliptic flow by roughly a factor of
two.

It is the purpose of this article to provide the basis for an
extension of our pQCD calculation toward the calculation of the
radiative energy loss. Some preliminary considerations have been
published in \cite{Gossiaux:2010yx}, where the calculation of
\cite{Gunion:1981qs} for the radiative cross section was extended to
the case of a collision implying one heavy
quark. In~\cite{Gossiaux:2010yx}, it is argued that for heavy quarks
of intermediate energy, those which constitute the bulk of the
production at RHIC and LHC, the gluon formation-time is strongly
reduced by mass effects, so that coherence effects can be discarded in
first approximation. In this respect, we offer a complementary
viewpoint to the works of~\cite{Dokshitzer:2001zm,
ASW,Zhang:2004,Djordjevic:2004,Zakharov:2004} where heavy
quarks are assumed to be ultrarelativistic and where the phase space
boundaries are not of primary importance. The same viewpoint will be
adopted in the present work in order to deduce and study the radiative
cross section that will be later implemented in our Monte Carlo
simulations in the same spirit as~\cite{Fochler:2013}. The colliding
light partons will be naturally considered as genuine dynamical
degrees of freedom -- see \cite{Djordjevic:2007at} as well -- and not
as fixed scattering centers, as it was the case in most of the
aforementioned works.

Starting out in section~\ref{sec:model} from the standard QCD
radiation matrix elements we calculate the gluon emission cross
section for the collisions of a heavy quark with a light quark. The
complexity of this result can be substantially reduced by realizing
that matrix elements can be regrouped into three gauge invariant
subgroups out of which one is dominating the energy loss. We then
establish that pQCD and scalar QCD (SQCD) give only slightly different
results as far as the energy loss of the heavy quark in a single
collision is concerned. Therefore we continue our calculation in the
SQCD approach which allows to compare our results with previous work
of Gunion and Bertsch for the light quark
sector~\cite{Gunion:1981qs}. We then discuss in
section~\ref{sec:gluon_distribution} the radiated gluon distribution
and in particular the ``dead cone'' effect, the suppression of almost
collinear gluon emission. This effect has been proposed a while ago by
Dokshitzer and Kharzeev~\cite{Dokshitzer:2001zm}. We show that the
emission of gluons with a small traverse momentum (with respect to the
direction of the incoming quark) is reduced but remains finite as soon
as this effect is calculated with gauge invariant matrix elements. In
section~\ref{sec:rad_cross_section} we calculate the fractional
radiative energy loss cross section $x\frac{d\sigma}{dx}$ as well as
its integral over $x$ entering the calculation of
the radiative energy loss $\frac{dE}{dz}$; we pay a particular
attention to the kinematic region for which $s-M^2\gg T^2$ but $s$ not
$\gg M^2$, relevant for production of heavy quarks at intermediate
$p_t$ in ultrarelativistic heavy ion collisions at RHIC and LHC. In
section~\ref{sec:gluon_mass} we extend the model by introducing a
finite gluon mass, as done in a number of phenomenological approaches
to study heavy ion collisions. We study in detail the influence of
such a mass for the radiative energy loss. In section
\ref{sec:energy_loss}, we then provide a comparison of radiative and
collisional energy loss. In an upcoming publication we will embed
these results into a numerical simulation of the radiative and
collisional energy loss using the hydrodynamical expansion scenario of
ref.~\cite{Werner:2010aa,Werner:2012xh}. Preliminary results for this
approach have been presented recently~\cite{gosssiaux:qnp}.

\section{Model}\label{sec:model}

\subsection{Matrix elements}

\begin{figure}
\begin{center}
\epsfig{file=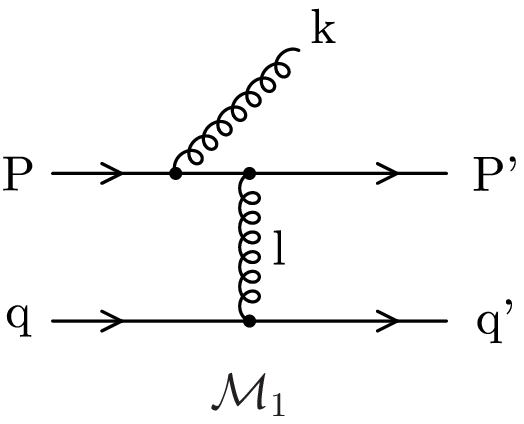,width=0.3\textwidth}
\epsfig{file=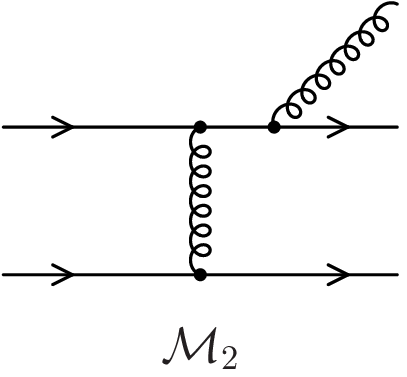,width=0.2\textwidth}\hspace{24pt}
\epsfig{file=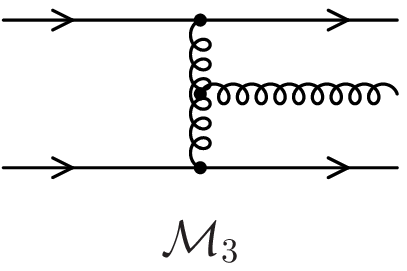,width=0.2\textwidth}\hspace{24pt}
\epsfig{file=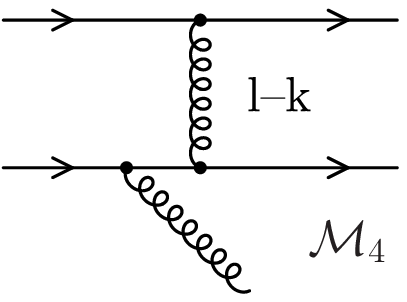,width=0.2\textwidth}\hspace{24pt}
\epsfig{file=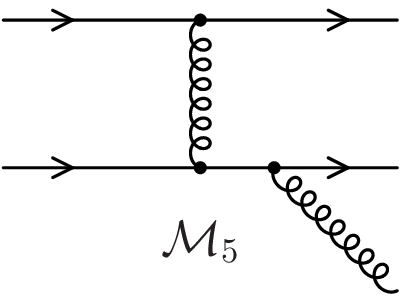,width=0.2\textwidth}
\end{center}
\caption{(Color online) The five matrix elements which contribute to
the gluon bremsstrahlung in a collision between a heavy quark of incoming
4-momentum $P$ and a light quark of incoming 4-momentum $q$.}
\label{dia}
\end{figure}

The starting point of our calculation are those five QCD
bremsstrahlung diagrams which are of the order of $g^3$ and describe
the creation of a gluon of 4-momentum $k$ in a collision between a heavy
quark $Q$ with mass $\mQ$ and incoming (outgoing) 4-momentum $\pa$
($\pone$) and a light quark $q$ taken as massless with incoming (outgoing)
4-momentum $\pb$ ($\pthree$) which is part of the plasma. They are shown
in Fig.~\ref{dia}. $\ell=\pb-\pthree$ is the 4-momentum transferred from
the light quark. The matrix elements are given in
Appendix~\ref{app:matrix_element} for completeness. We found that the
quark spin is inessential. Considering scalar quarks is then
sufficient and we give here $\mathcal{M}_1$, $\mathcal{M}_2$ and
$\mathcal{M}_3$ in scalar QCD (SQCD, whose Feynman rules can be found in~\cite{IZ,Meggiolaro:1995cu})
\begin{eqnarray}
\mathcal{M}_1&=& C_1 \tilde{\mathcal{M}}_1 = g^3 C_1 \
\frac{(\pb+\pthree)^\mu}{(\pthree-\pb)^2}D_{\mu\nu}[\pthree-\pb]
\left\{ \frac{(\pa-k+\pone)^\nu
(2\pa-k)\cdot\epsilon}{(\pa-k)^2-\mQsq} - \epsilon^\nu \right\},\nonumber\\
\mathcal{M}_2&=& C_2 \tilde{\mathcal{M}}_2= g^3 C_2\ 
\frac{(\pb+\pthree)^\mu}{(\pthree-\pb)^2}D_{\mu\nu}[\pthree-\pb]
\left\{ \frac{(\pa+\pone+k)^\nu
(2\pone+k)\cdot\epsilon}{(\pone+k)^2-\mQsq}-\epsilon^\nu \right\},\nonumber\\
\mathcal{M}_3&=&C_3 \tilde{\mathcal{M}}_3= g^3 C_3
\frac{(\pb+\pthree)^\mu(\pa+\pone)^\nu}{(\pthree-\pb)^2(\pone-\pa)^2}
\,D_{\mu\mu'}[\pthree-\pb]D_{\nu\nu'}[\pone-\pa]\nonumber \\
&\times&
\left\{g^{\mu'\nu'}(\pa-\pone+\pthree-\pb)^\sigma+
g^{\nu'\sigma}(\pone-\pa-k)^{\mu'}+g^{\sigma\mu'}(\pb-\pthree+k)^{\nu'}\right\}
\epsilon_\sigma\,. 
\label{eq_def_matrix_element}
\end{eqnarray}
In $\mathcal{M}_1$ and $\mathcal{M}_2$ the $\epsilon^\nu$-term comes
from the extra diagram in SQCD where the emitted gluon is
attached to the upper quark-exchanged-gluon vertex. We 
work in light cone gauge, $A\cdot n=0$ with $n$  being a fixed light-like vector, for which we find 
$D_{\mu\nu}[\ell]=-g_{\mu\nu} +\frac{\ell_\mu n_\nu + \ell_\nu
n_\mu}{\ell\cdot n}$. The color matrix elements are
\begin{equation}\label{eq:color_factor}
C_1=(T^b T^a)(T^b),\quad
C_2=(T^a T^b)(T^b),\quad
C_3=if_{abc} (T^c)(T^b),
\end{equation}
where the color matrix in the first (second) bracket is that of the
heavy (light) quark. The two remaining matrix elements $\mathcal{M}_4$
and $\mathcal{M}_5$ can be obtained from $\mathcal{M}_1$ and
$\mathcal{M}_2$ by exchanging heavy and light quark momenta
($\pa\leftrightarrow\pb$, $\pone\leftrightarrow\pthree$) and color
labels.

The commutation relation
\begin{equation}
T^bT^a=T^aT^b-if_{abc}T^c,
\end{equation}
allows for regrouping the five matrix elements into three combinations,
each of them being gauge invariant:
\begin{eqnarray}\label{eq:color_decomposition}
\mathcal{M}^{\rm SQED}_{Q} &=& C_2 (\tilde{\mathcal{M}}_1+\tilde{\mathcal{M}}_2),\nonumber\\
\mathcal{M}^{\rm SQED}_{q} &=& C_5 (\tilde{\mathcal{M}}_4+\tilde{\mathcal{M}}_5),\nonumber\\
\mathcal{M}^{\rm SQCD} &=& C_3(\tilde{\mathcal{M}}_3-\tilde{\mathcal{M}}_1+\tilde{\mathcal{M}}_4),
\end{eqnarray}
where the $\tilde{\mathcal{M}}_i$ are defined in eq.~\ref{eq_def_matrix_element}.
$Q$ ($q$) marks the emission of the gluon from the heavy (light) quark
line. $C_5$ is obtained from $C_2$ by exchanging the heavy quark and
light quark in eq.~\ref{eq:color_factor}. The combination of diagrams
labeled as SQED are the bremsstrahlung matrix elements already present
in scalar Quantum Electrodynamics (SQED) whereas the amplitude labeled
SQCD is a genuine matrix element of Quantum Chromo
Dynamics. $\mathcal{M}^{\rm SQCD}$ is the main objet of interest
here. It dominates the energy loss of heavy quarks, as we will
show. In passing we mention that the decomposition of the five
amplitudes into gauge invariant subgroups of diagrams is not
unique. Beside the decomposition shown in
eq.~\ref{eq:color_decomposition} one can find a decomposition into
commuting and anticommuting color operators. Such a decomposition has
the advantage that the interference term disappears but the
inconvenience of lengthy expressions.

\subsection{Differential cross section at finite energy (model I)}
\label{sub:diff_x_sec}

It is convenient to specify the kinematics using a Sudakov
decomposition of momenta. Pick $\pb$ as a light-like momentum 
(here chosen as the 4-momentum of the massless light quark) and
choose $p$ such that $p^2=0$ and $\pa=p+b\,\pb$. From $\pa^2=\mQsq$
and $s=(\pa+\pb)^2$, it follows that
\begin{equation}
\pa=p+\frac{\mQsq}{2\,p\cdot\pb}q,\quad s=\mQsq+2\,p\cdot\pb.
\end{equation}
The emitted gluon 4-momentum thus reads as
\begin{equation}
k=x\,p+k_q\,q+\kt,\quad
\mathrm{with}\quad k_q=\frac{\vec{k}_t^2}{2x p\cdot\pb}.
\end{equation}
In this form, it is clear that the momentum fraction $x$ is a Lorentz
invariant ($x=k\cdot\pb/p\cdot\pb$) and that $\kt$ is a space-like
4-vector which is transverse to both $p$ and $\pb$.
Thus, it is equivalent to $\vec{k}_t\in\mathbb{R}^2$,
with a norm $\kt^2=-\vkt^2$. Writing $\ell=\ell_p\,p+\ell_q\,q+\lt$, the
set of independent variables can be chosen as $s$, $x$, the magnitude
of $\vkt$, that of $\vlt$ and $\vkt\cdot\vlt$, or the angle
$\phi_{k_t}$ between $\vkt$ and $\vlt$. $\mathcal{M}^{\rm SQCD}$ then reads as follows:
\begin{equation}\label{eq:MSQCD}
\mathcal{M}^{\rm SQCD}=
g\,C_3\,\left(\frac{-4\,g^2\,\pa\cdot\pb}{\ell^2}\right)
\left(\frac{(2(1-x)-x')\,\vec{\epsilon}_t\cdot\vkt}{\vkt^2+x^2\mQsq}
-\frac{2(1-x-x')\,\vec{\epsilon}_t\cdot(\vkt-\vlt)}{(\vkt-\vlt)^2
  +(x+x')^2\mQsq}\right).
\end{equation}
In order to keep the matrix element in a compact form, the extra
variable $x' \equiv -\ell_p=-\ell^2/(2p\cdot q)$ is used in addition to
the set of independent kinematical variables $(x,\vkt,\vlt)$. 
The appearance of $x'$ and its deduction are
explained in the Appendix~\ref{app:phase_space}. Obtaining
eq.~\ref{eq:MSQCD} is straightforward, noticing that the occurrence of
$x'$ in the numerator partly comes from the identity
$\frac{-2\,\pa\cdot\pb}{\ell^2} \times x'=1$. The interest to factorize out
$\frac{-2\,\pa\cdot\pb}{\ell^2}$ becomes clear at high
energy as will be discussed shortly.

As it stands the matrix element is infrared sensitive. 
For a $qQ$ scattering taking place inside a QCD medium at finite temperature, 
the gluon propagator acquires finite electric and magnetic thermal masses \cite{Silva:2013} $m_D \sim g T$ and $m_{\rm mag}\sim g^2T$ which are usually interpreted in terms of screening effects. This screening prevents the cross section from being sensitive to $\sqrt{-\ell^2}$ smaller than the gluon mass, which acts therefore as a typical momentum transfer. The prescription for regularization that we adopt is to multiply the amplitude $\mathcal M$ by $\vlt^2/(\vlt^2+\mD^2)$, with $\mD\sim g T$. There are other
propositions, to use hard thermal loop calculations \cite{Braaten:1991jj,Djordjevic:2007at} or to introduce a self consistent temperature dependent Debye mass \cite{Peshier:2007}.

After squaring the matrix element, summing over the transverse
polarizations, and making use of the phase space integral derived in
Appendix~\ref{app:phase_space}, one obtains, for the gluon emission cross section:
\begin{equation}
\frac{d\sigma^{Qq\rightarrow Qgq}}{dx d^2\kt d^2\lt} =
\frac{1}{2(s-\mQsq)}|\mathcal{M}|^2 \frac{1}{4(2\pi)^5\,\sqrt{\Delta}}
\Theta(\Delta)\,,
\label{eq:xsec}
\end{equation}
with 
\begin{equation}
\Delta=\left(x(1-x)\,s-x\,\mQsq-\vkt^2+2x\,\vkt\cdot\vlt\right)^2
-4x(1-x)\,\vlt^2\,(x\,s-\vkt^2).
\label{eq:def_Delta}
\end{equation}

The evaluation of $\frac{d\sigma^{Qq\rightarrow Qgq}}{dx d^2\kt
d^2\lt}$, eq.~\ref{eq:xsec}, with $\mathcal{M}=\mathcal{M}^{\rm SQCD}$,
eq.~\ref{eq:MSQCD}, gives what we call the finite energy cross
section, model~I. With the presence of $x'$ and its somewhat
complicated dependence on the other variables the full result does not
allow for an easy physical discussion at all energies. In
section~\ref{sub:wloss} we present the numerical results in which the
full matrix element, eq.~\ref{eq:MSQCD}, is used and the subsequent
integration over the phase space variables has been done by a Monte
Carlo method. This allows for taking into account the boundaries of the integration in a very convenient way.

The physics becomes more transparent when we go to the high energy
limit where subleading terms in $\sqrt{s}$ are neglected. This limit
has also the advantage that the expressions for the differential as
well as for the various integrated cross sections become very
compact. One of the aims of the present study is to investigate the
accuracy of the high-energy approximation with respect to the full
result.

\subsection{High-energy approximation and model II}

To elucidate the physics of the gluon emission we specify the
high-energy regime of interest. Assuming that $s\approx
2\pa\cdot\pb$, i.e., $\mQsq\ll s$, our discussion parallels that
of Gunion and Bertsch~\cite{Gunion:1981qs}. At the end of the section
we examine more deeply the interplay between $\mQsq$ and $s$ in order
to specify what happens if $\mQsq\ll s$ is not fulfilled. However,
this discussion is easier to carry out \textit{a posteriori}.

At high energy there is room for radiation in a wide rapidity
interval~\cite{Gunion:1981qs}. The central region, $x\ll 1$ and 
$k_q\equiv\frac{\vkt^2}{2x p\cdot\pb}\ll 1$, is driven by 
$\mathcal{M}^\mathrm{SQCD}$, which thus
forms the bulk of radiation (see~\cite{Gunion:1981qs} and also the
discussion in section~\ref{sec:gluon_distribution}).
$\mathcal{M}^\mathrm{SQED}_Q$ and $\mathcal{M}^\mathrm{SQED}_q$ become
competitive respectively at large $x$ and at large $k_q$
(corresponding to very small $x$), but are otherwise suppressed. In
section~\ref{sec:gluon_distribution}, it is shown that the important
region of phase space for radiation is $\vkt^2\lesssim|\ell^2|$. This
is a consequence of the $1/(\vkt^2)^2$ behavior of the differential
cross section, eq.~\ref{eq:xsec}, with the matrix element,
eq.~\ref{eq:MSQCD}, at large $\vkt^2$ for fixed $t\equiv\ell^2$. In
addition the $1/t^2$ suppression of the differential cross section at
large $|t|$ makes the large $|t|$ region essentially irrelevant.

We therefore define the large $s$ domain as
\begin{equation}\label{eq:high-energy}
s,x\,s\gg |\ell^2|, \vkt^2.
\end{equation}
It encompasses the rapidity interval between the fragmentation regions
of heavy and light quarks where $x$ is small but not that small that
it approaches the light quark fragmentation regime
$x\sim\vkt^2/s$. The latter region will be discarded from the
analysis. Eq.~\ref{eq:high-energy} also covers the heavy quark
fragmentation region, where $x$ is finite, $x\sim 1$, allowing us to
quantify the relative importance of $\mathcal{M}^\mathrm{SQCD}$ and
$\mathcal{M}^\mathrm{SQED}_Q$, see Appendix~\ref{app:qed_vs_qcd}.

In leading order of $s$ all squares of the matrix elements,
$|\mathcal{M}|^2$, factorize
\begin{equation}\label{eq:pgdef}
|\mathcal{M}|^2 =16\pi^3x(1-x)\,|\mathcal{M}_{\rm el}(s,\vlt)|^2
 P_g(x,\vkt,\vlt),
\end{equation}
with $|\mathcal{M}_{\rm el}(s,\vlt)|^2 =\frac{C_F}{2N_c}
g^4\frac{4s^2}{(\vlt^2+\mu^2)^2}$ being the regularized matrix element
squared for the elastic cross section at high energy
($C_F/(2N_c)=2/9$). As a consequence the differential cross section
can be written as
\begin{equation}
\frac{d\sigma^{Qq\rightarrow Qgq}}{dx d^2\kt d^2\lt}
\underset{s\gg M^2}{\to}
\frac{d\sigma_{\rm el}}{d^2\lt}\, P_g(x,\vkt,\vlt),
\label{dsigma_high_energy}
\end{equation}
with $\frac{d\sigma_{\rm el}}{d^2\lt}\rightarrow\frac{8\alpha_s^2}{9(\vlt^2+\mu^2)^2}$. 
We mention that the spin averaged square of the QCD matrix element is
the sum of $|\mathcal{M}^{\rm SQCD}|^2$ which is the squared matrix
element for the same bremsstrahlung process in SQCD and a correction
term which is negligible at small $x$, the dominating region of the
gluon emission:
\begin{equation}\label{eq:M2QCD}
\frac{1}{4}\sum_{\rm spin}|\mathcal{M}^{\rm QCD}|^2
=|\mathcal{M}^{\rm SQCD}|^2+\frac{x^2}{2(1-x)}\frac{\vlt^2}{(\vkt^2+x^2\mQsq)
\left((\vkt-\vlt)^2+x^2\mQsq\right)}C_A 16\pi \alphas (1-x)^2
|\mathcal{M}_{\rm el}|^2.
\end{equation}
The color factor is $C_A=3$. Thus at small $x$, as we will see, the
dominant region for the energy loss as well as for the gluon emission,
spinor QCD can be well approximated by scalar QCD and we can profit
from the fact that $|\mathcal{M}^{\rm SQCD}|^2$ has in the large energy
limit a very simple form:
\begin{equation}
|\mathcal{M}^{\rm SQCD}|^2= C_A 16\pi \alphas (1-x)^2
|\mathcal{M}_{\rm el}|^2\left(\frac{\vkt}{\vkt^2+x^2\mQsq}
-\frac{\vkt-\vlt}{(\vkt-\vlt)^2 +x^2\mQsq}\right)^2,
\label{eq:M2SQCD}
\end{equation}
leading to
\begin{equation}
P_g(x,\vkt,\vlt;M)=\frac{C_A\alphas}{\pi^2}\frac{1-x}{x}
\left(\frac{\vkt}{\vkt^2+x^2\mQsq}
-\frac{\vkt-\vlt}{(\vkt-\vlt)^2 +x^2\mQsq}\right)^2.
\label{eq:gluon_distribution}
\end{equation}
In light-cone gauge with fixing gauge vector $n=\pb$, the first term
in the bracket describes the emission from the incoming heavy quark
line and the second term the emission from the gluon. This shows that
in this gauge and away from the light quark fragmentation region the
matrix element for the emission from the light quark does not
contribute. In Sect.~\ref{sub:wloss} the comparison between the full result,
model~I, and the high-energy approximation will allow for a
quantitative judgement of the relevance of the latter in the
phenomenologically accessible range of $\sqrt{s}$.

The high-energy approximation eq.~\ref{eq:M2SQCD} is easily obtained by setting  $x'\to 0$ in eq.~\ref{eq:MSQCD} .
Using this result in eq.~\ref{eq:xsec} and approximating $\sqrt{\Delta}\approx x(1-x)\,s$,
$s-\mQsq\approx s$ and $-t\approx\vlt^2$, but keeping the exact phase
space boundaries eqs.~\ref{eq:boundaries}, gives an approximation,
referred to as model~II, that incorporates part of the finite energy
corrections easy to implement:
\begin{equation}
\frac{d\sigma^{Qq\rightarrow Qgq}_{II}}{dx d^2\kt d^2\lt}=
\frac{d\sigma_{\rm el}}{d^2\lt}\, P_g(x,\vkt,\vlt)\Theta(\Delta)\,.
\label{def:model2}
\end{equation}
Model II is situated midway between the full calculation, model I, and the high-energy approximation eq.~\ref{dsigma_high_energy}.

We observe that the high-energy approximation does not necessarily
require $s$ to be much bigger than $\mQsq$ in $\mathcal{M}^{\rm
SQCD}$. Only $x'\ll x$ and $x'\ll 1-x$ are mandatory. The first
condition writes $-t\ll x(s-\mQsq)$, which can be fulfilled for
moderate $t$ even if $s$ is not very large with respect to $\mQsq$. In
such a circumstance, one power of $2\,\pa\cdot\pb= s-\mQsq$ in
$|\mathcal{M}^{\rm SQCD}|^2$ cancels out with the same factor in the
denominator of eq.~\ref{eq:xsec} while the second power cancels out
exactly at small $x$ where $\sqrt{\Delta}\approx x\,(s-\mQsq)$. Thus,
imposing both $x'\ll x$ and $k_q\ll 1$, corresponding to $\vkt^2\ll
x\,(s-\mQsq)$, eq.~\ref{eq:high-energy} is replaced by
\begin{equation}
s-\mQsq,\ x\,(s-\mQsq)\gg|\ell^2|,\vkt^2.
\label{eq:hypsmalltransverse}
\end{equation}

In the case of massless quarks, eq. \ref{eq:M2SQCD} is identical with 
the matrix elements of Gunion and Bertsch (GB) of ref.~\cite{Gunion:1981qs}. Their discussion
at the amplitude level is carried through within SQCD at both small $x$ and
finite $x$. In spinor QCD, even for massless quarks, we were able to
compute finite $x$ correction only at the squared amplitude level. At
$\mQ=0$, we observe the factorization between transverse ($\vkt$,
$\vlt$) and longitudinal ($x$) dependence, and for the latter a factor
$2(1-x)$ in SQCD and $2(1-x)+x^2$ in spinor QCD which is reminiscent
of the quark splitting functions in SQCD and QCD.

Similarly we derive the QED-like terms. $|\mathcal{M}^{\rm QED}_{Q}|^2$
can be directly drawn from eqs~\ref{eq:M2QCD} and~\ref{eq:M2SQCD} by
changing $\vlt\to x\vlt$ and the color factor $C_A\to C_F=4/3$. 
At small $x$, one gets $|\mathcal{M}^{\rm QED}_{Q}|^2=
|\mathcal{M}^{\rm SQED}_{Q}|^2+\mathcal{O}(x^2)$ with
\begin{equation}
\label{eq:M2SQED}
|\mathcal{M}^{\rm SQED}_{Q}|^2= C_F \ 16\pi \alphas(1-x)^2
|\mathcal{M}_{\rm el}|^2\left(\frac{\vkt}{\vkt^2+x^2\mQsq}
-\frac{\vkt-x\vlt}{(\vkt-x\vlt)^2+x^2\mQsq}\right)^2.
\end{equation}

\section{Gluon distribution in the high energy approximation}
\label{sec:gluon_distribution}

In this section, we study the gluon distribution $P_g$ as a function
of $\kt$ (from now on, $\kt$ will refer to $|\vec{k}_t|$ and $\lt$ to 
$|\vec{l}_t|$). $P_g$ exhibits the well-known bremsstrahlung
phenomenology. When $\lt$ and $x\mQ$ are incommensurate, there are two
extreme regimes: the hard scattering regime, $\lt\gg x\mQ$, and the
soft scattering regime, $\lt\ll x\mQ$. Inspection of
eq.~\ref{eq:M2SQCD} shows that the important region for radiation is
that of intermediate $\kt$ since $P_g$ remains finite at small $\kt$
and $P_g\propto \kt^{-4}$ at large $\kt$. Thus, for a hard scattering
we find
\begin{equation}\label{eq:hard_scatt_dist}
P_g\propto\frac{\vkt^2}{(\vkt^2+x^2\mQsq)^2},
\end{equation}
assuming $\frac{x^2\mQsq}{\lt}\ll\kt\ll\lt$. In the hard scattering
regime, the radiation is logarithmically enhanced for $\kt\gg x\mQ$
and there is a dead cone for $\kt < x\mQ$. There the ratio between the
gluon distribution function for the massive and massless cases reads
\begin{equation}\label{eq:KHQ}
K_\mathrm{HQ}=\frac{P_g(x,\vkt,\vlt;\mQ)}{P_g(x,\vkt,\vlt;0)}
=\frac{1}{\left(1+\frac{x^2\mQsq}{\vkt^2}\right)^2}.
\end{equation}
The situation of hard scatterings on the medium is implicitly assumed
in the analysis of Ref.~\cite{Dokshitzer:2001zm}. 
For soft scattering, $\lt\ll x\mQ$, there is a strong
interference between both factors in the bracket of
eq.~\ref{eq:M2SQCD} and no room for large (i.e. log-enhanced)
radiation.

Considering charm or bottom quarks, $\mQ=1.3$, 4.5~GeV, in a medium
characterized by $\mD=0.4$~GeV, both regimes are
encountered in the $x,\lt$ plane and we now study quantitatively the
resulting $\kt$ dependence.

\subsection{Integration over $\angle(\vlt,\vkt)$}

In the high energy limit the integral over $\phi_{k_t}$ can be
performed analytically over $[0,2\pi]$:
\begin{eqnarray}
\mathcal{I}(x,\kt,\lt;\mQ)&=& \int d\phi_{k_t} P_g(x,\vkt,\vlt)
=\frac{2 C_A\alphas}{\pi}\frac{1-x}{x}
\left[
\frac{\vlt^2+2x^2\mQsq}{\left(\vkt^2+x^2\mQsq\right)
\left(\left(\vkt^2+x^2\mQsq+\vlt^2\right)^2
-4\vkt^2\vlt^2\right)^{1/2}}
\right.\nonumber\\
&&\left.-\frac{x^2\mQsq}{(\vkt^2+x^2\mQsq)^2}
-\frac{x^2\mQsq(\vkt^2+x^2\mQsq+\vlt^2)}
{\left(\left(\vkt^2+x^2\mQsq+\vlt^2\right)^2
-4\vkt^2\vlt^2\right)^{3/2}}\right],
\label{gbia}
\end{eqnarray}
and one finds for zero mass quarks as well the GB result:
\begin{equation}
\mathcal{I}(x,\kt,\lt;\mQ=0)=\frac{2 C_A\alphas}{\pi}\frac{1-x}{x}
\frac{\vlt^2}{\vkt^2 |\vkt^2-\vlt^2|}.
\label{IGB}
\end{equation}
Again, the corresponding SQED term can be obtained by replacing
$\vlt\to x\vlt$ and $C_A\to C_F$.

\begin{figure}[H]
\begin{center}
\epsfig{file=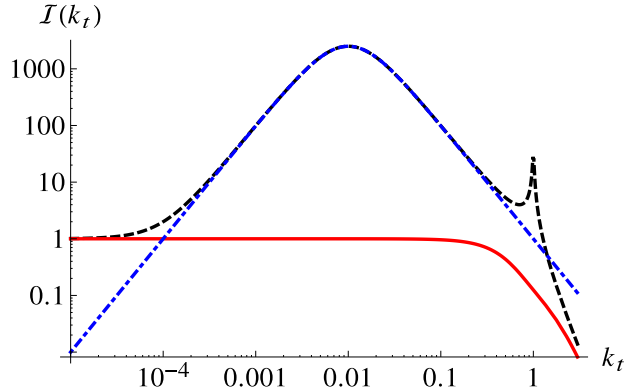,width=0.45\textwidth}
\end{center}
\caption{(Color online) $\kt$ distribution of gluons. The full red line and the dashed black line correspond to 
$\mathcal{I}(x,\kt,t;\mQ)$, eq.~\ref{gbia}, with $\lt=0.01,x\mQ=1$
and $\lt=1,x\mQ=0.01$, respectively, while the dash-dotted blue line 
corresponds to the dead cone distribution, eq.~\ref{eq:hard_scatt_dist}, 
with $x\mQ=0.01$.}
\label{fig:gluon_kt}
\end{figure}

Fig.~\ref{fig:gluon_kt} shows the behaviour of $\mathcal I$ for
$\lt=1$, $x\mQ=0.01$ (dashed black line) and $\lt=0.01$, $x\mQ=1$
(full red line), chosen to make visible the aforementioned hard and
soft scattering regimes. In Fig.~\ref{fig:gluon_kt} $\mathcal{I}$ is 
divided by its value at $\kt=0$:
\begin{equation}\label{eq:kt_to_0}
\mathcal{I}(x,0,\lt;\mQ)=\frac{2 C_A\alphas}{\pi}\frac{1-x}{x}
\frac{\vlt^2}{(\vlt^2+x^2\mQsq)^2}.
\end{equation}
For comparison, the ``dead cone'' distribution
eq.~\ref{eq:hard_scatt_dist} is also plotted for $x\mQ=0.01$
(dash-dotted blue line).

We observe that for hard scattering $\mathcal{I}(x,|\vkt|,\lt;\mQ)$
falls off $\propto 1/|\vkt|^4$ for large $\kt$. This can be directly
read off from eq.~\ref{eq:M2SQCD} setting $\mQ=0$ which leads us back
to the Gunion-Bertsch behavior, eq.~\ref{IGB}. For $\kt$ in between
$(x\mQ)^2/\lt$ and $\lt$, $\mathcal{I}$ behaves as indicated in
eq.~\ref{eq:hard_scatt_dist}, which corresponds to dropping the second
term in the bracket of eq.~\ref{eq:M2SQCD}, hence we observe the
absence of interference in this range: on the plot, the dash-dotted
line is on top of the dashed line. As anticipated, we observe a
log-enhanced radiation $\mathcal{I}\propto 1/\vkt^2$ for
$x\mQ<\kt<\lt$ and a dead cone suppression for
$(x\mQ)^2/\lt<\kt<x\mQ$, as well as a maximum at
$k_t\sim x M$. The second region of enhanced radiation on
Fig.~\ref{fig:gluon_kt}, visible around $\kt=\lt$ , is due to the
second term in the bracket of eq.~\ref{eq:M2SQCD}.

For soft scattering, keeping only terms of at most
$\mathcal{O}(\vlt^2)$, we find
\begin{equation}\label{eq:soft_scatt_dist}
\mathcal{I}(x,\kt,\lt;\mQ)=\frac{2 C_A\alphas}{\pi}\frac{1-x}{x}
\frac{\vlt^2}{(\vkt^2+x^2\mQsq)^2}\times
\frac{|\vkt|^4+x^4\mQ^4}{(\vkt^2+x^2\mQsq)^2}.
\end{equation}
This behavior results from a strong interference between the two terms
in the bracket of eq.~\ref{eq:M2SQCD}, since both terms can be large
but their difference is small in this regime. The full line shows the
direct transition from a constant value at small $\kt$ to a  $1/|\vkt|^4$ dependence for
large $\kt$. This behavior is easily obtained from the approximation in
eq.~\ref{eq:soft_scatt_dist}, noticing that the last factor is 1 in
both limits.

\subsection{Gluon emission in $x-k_t$ space and the dead cone effect}
\label{sub:small_kt}

\begin{figure}[H]
\begin{center}
\epsfig{file=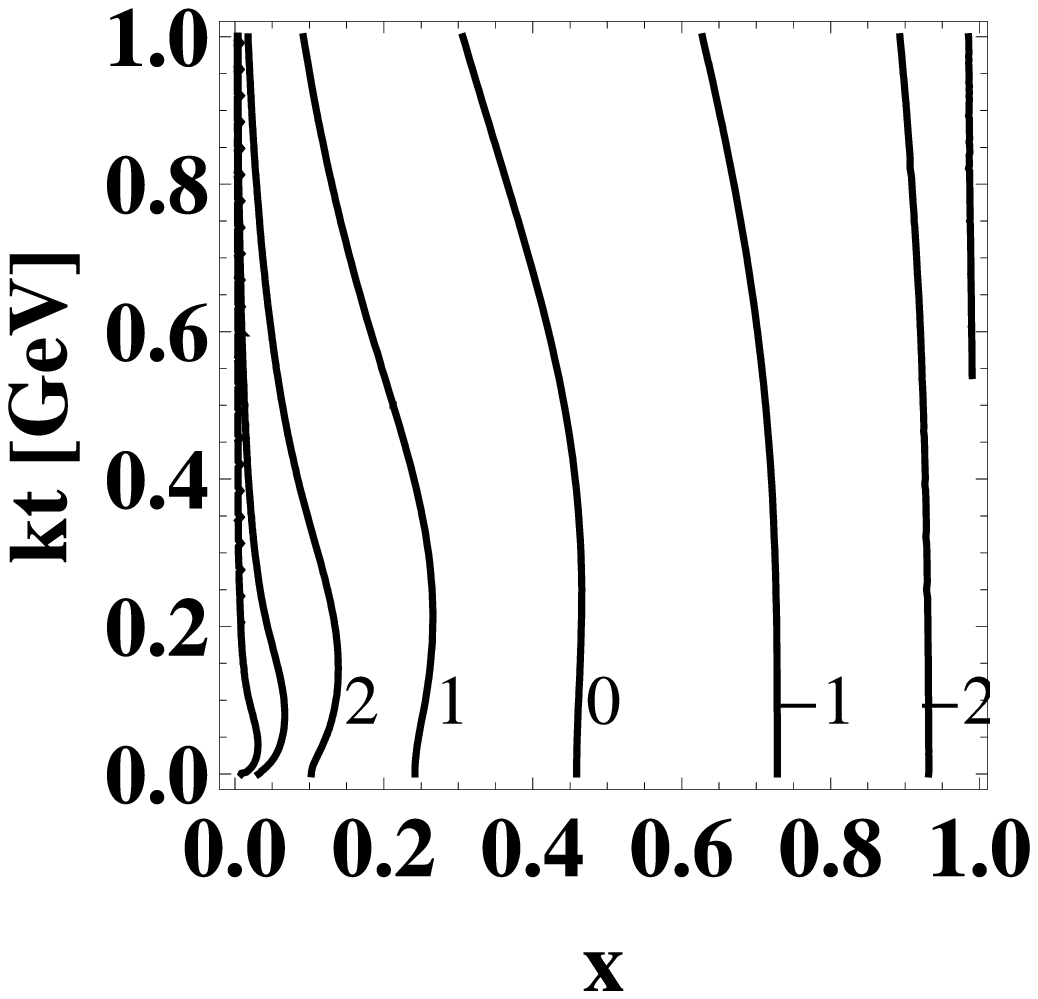,width=0.4\textwidth}
\epsfig{file=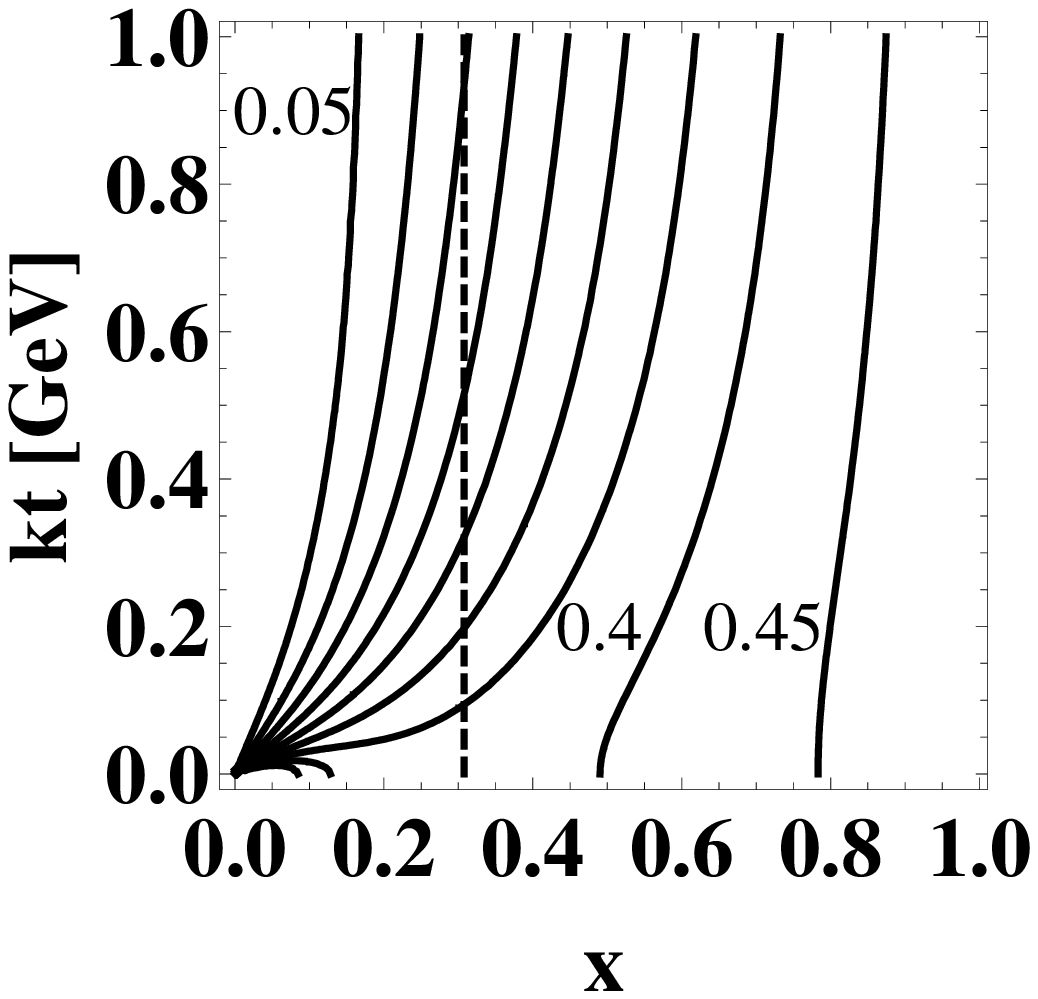,width=0.4\textwidth}
\end{center}
\caption{Left: Contour plot of the $x$ and $\kt$ distribution 
of gluons emitted from charm quarks, $\bar P_g(x,\kt;\mQ=1.3~\mathrm{GeV})$,
  in logarithmic scale. Right: Contour plot of the relative contribution of
  the QED-like cross section to the total gluon emission cross section
  for $\mQ=1.3~\mathrm{GeV}$ as a function of $x$ and $\kt$.
  The infrared regulator is chosen as $\mD=0.4$~GeV. The dashed line at
  $x=\mu/M$ indicates the transition from hard scattering regime 
  ($x\lesssim \mu/M$) to soft scattering regime ($x\gtrsim \mu/M$).}
\label{deadf}
\end{figure}

One step further can be made by averaging over the elastic cross
section, defining
\begin{equation}\label{eq:av_gluon_dist}
{\bar P}_g(x,\kt;\mQ)=\frac{\int d^2\vlt \frac{1}{(\vlt^2+\mD^2)^2}
P_g(x,\vkt,\vlt)}%
{\int d^2\vlt \frac{1}{(\vlt^2+\mD^2)^2}},
\end{equation}
with an infrared regularization for the elastic cross section as
discussed above. Let us notice that due to the fast decrease of the differential elastic 
cross section with $|\vec{l}_t|$, the gross features of ${\bar P}_g$ can be obtained from those of $P_g$ -- or more precisely from those of ${\cal I}/(2\pi)$ in eq.~\ref{gbia} -- by substituting  $\mD$ for $|\vec{l}_t|$.
Contours of the distribution of gluons emitted from
a charm quark as a function of $x$ and $\kt$, ${\bar P}_g(x,\kt;\mQ)$,
are shown in fig.~\ref{deadf}(left). We have illustrated 
the case $\mu<M$ which includes both hard and soft scattering regimes.
We see that the radiation is
concentrated at small $x$ (in the hard scattering regime) and small $\kt$ - values. 

The QED-like term, eq.~\ref{eq:M2SQED}, contributes only little to the
overall radiation, as can be seen in Fig.~\ref{deadf}(right) which
shows contours of the ratio ${\bar P}_g^{\rm SQED}/{\bar P}_g^{tot}$
as a function of $x$ and $\kt$ for $\mQ=1.3$~GeV. ${\bar P}_g^{tot}$
is built from $|\mathcal{M}^{\rm SQCD}+\mathcal{M}_Q^{\rm SQED}|^2$
whereas ${\bar P}_g^{\rm SQED}$ is built from $|\mathcal{M}_Q^{\rm SQED}|^2$.

In the regime of hard scattering, at small 
$x\lesssim \mD/M$, the QED-like contribution is marginal for almost
all $k_t$. The ratio becomes sizable (property A) for very small $\kt$ only, 
$\kt < \frac{x^2 M^2}{\langle l_t \rangle} \approx 
\frac{x^2 M^2}{\mD} < x M$, corresponding to large rapidities~\cite{Gunion:1981qs}. The SQED contribution can even become
the dominant one (property B) for $x<\mD^2/\mQsq$ and $\kt < x \langle l_t \rangle 
\approx x\mD < x M$. In both cases however, those regions of phase
phase space are very limited in comparison to the range
$\kt\in[x \mQ,\mD]$ where the QCD radiation is large. 
The ratio becomes sizable (property C) also in the soft scattering 
regime (at large $x$), where the radiation is weak.
A more detailed analysis is given in Appendix~\ref{app:qed_vs_qcd}. 
The second QED-like term, $|\mathcal{M}^{\rm SQED}_{q}|^2$, describing the
gluon emission from the light quark, is irrelevant at high-energy away
from the light-quark fragmentation region.

\begin{figure}[H]
\begin{center}
\epsfig{file=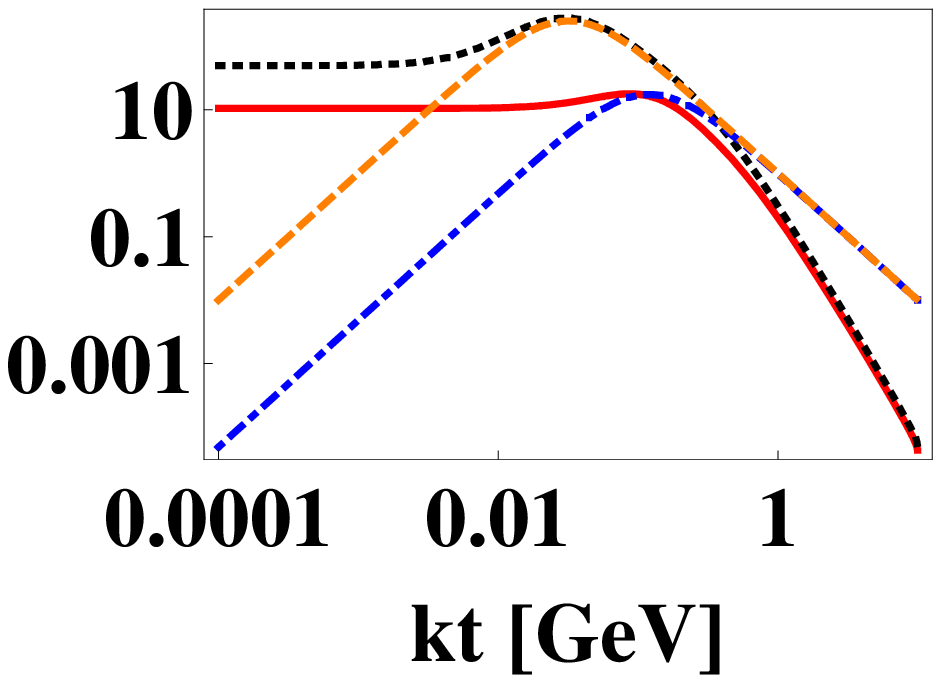,width=0.45\textwidth}
\epsfig{file=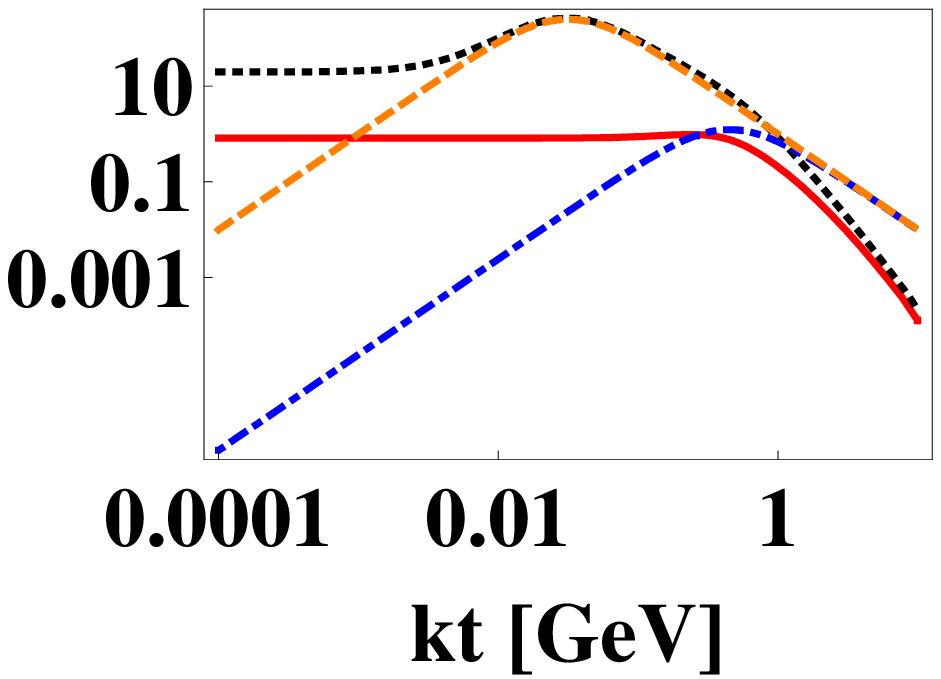,width=0.45\textwidth}
\end{center}
\caption{(Color online) The suppression of gluon radiation of quarks,
$\bar P_g(x=0.1,\kt ,\mQ)$, in our approach (red full line and black
dotted line for the emission from a heavy quark and of a quark with
mass $\mQ=0.1$~GeV, respectively) as compared to the hard scattering
approach, eq.~\ref{eq:hard_scatt_dist}, (blue dash-dotted line and
orange dashed line for the emission from heavy and $\mQ=0.1$~GeV
quarks, respectively). On the left hand side we display the results
for charm quarks, $\mQ=1.3$~GeV, and on the right hand side for bottom
quarks, $\mQ=4.5$~GeV.}
\label{deadf2}
\end{figure}

Fig.~\ref{deadf2} compares our results, eq.~\ref{eq:av_gluon_dist},
including the squared matrix element given by eq.~\ref{eq:M2SQCD},
with the dead cone approximation eq.~\ref{eq:hard_scatt_dist}.  The
red full line shows on the left (right) hand side the $\kt$
distribution of gluons at $x=.1$, ${\bar P}_g(x=0.1,\kt;\mQ)$, emitted
from charm (bottom) quarks, and the black dashed line shows the
distribution of gluons emitted from light quarks with
$\mQ=0.1$~GeV. These curves are compared with the results of the hard
scattering approach (blue dash-dotted line for the heavy quark and
orange dashed line for the light quarks). The main features, discussed
in the last section for a fixed $\lt$, are still visible after the
averaging over elastic cross section, $\lt$ being replaced by
$\mD$. For the light quark, and the chosen $x$-value, $x\mQ$ is much
smaller than $\mD$ and the characteristics of the hard scattering
regime are visible. In particular, we observe a strong radiation
window at intermediate $\kt\lesssim\mD$ that is fairly well reproduced
by the dead cone approximation. For the bottom quark, for which $x\mQ$
is comparable to $\mD$, the trend is typical of the soft scattering
regime, with nowhere a match with the dead cone approximation.

The suppression of the radiation from heavy quarks as compared to that
from light quarks at small $\kt$ can be quantified by a suppression
factor $\bar{K}$
\begin{equation}
\bar{K}(x,\kt;\mQ)= 
{\bar P}_g(x,\kt;\mQ)/{\bar P}_g(x,\kt;\mQ=0.1\ \mathrm{GeV}).
\end{equation}

\begin{figure}[H]
\begin{center}
\epsfig{file=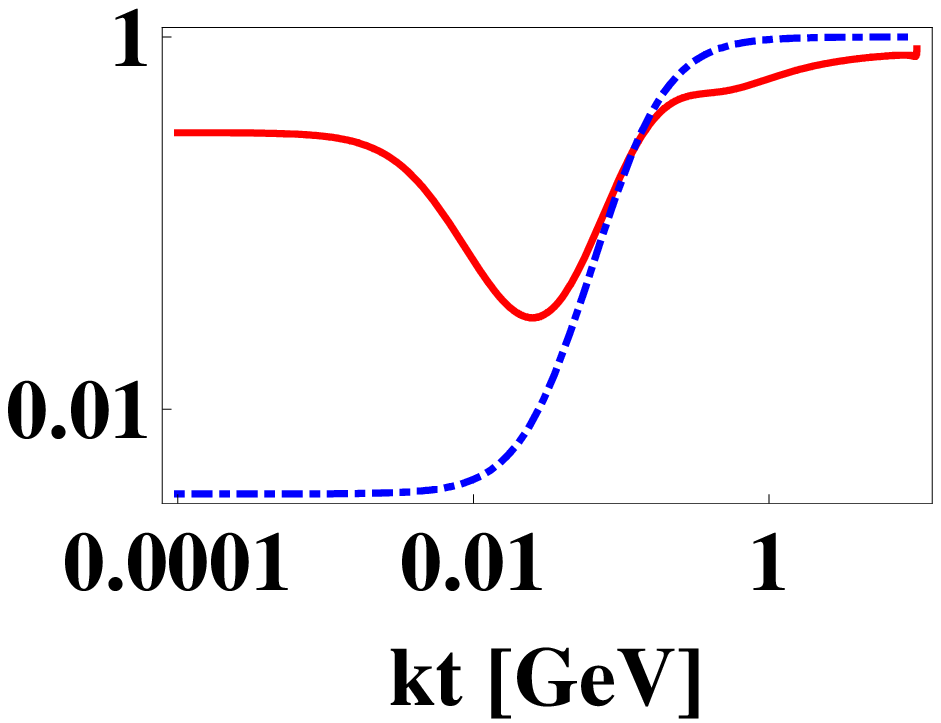,width=0.45\textwidth}
\epsfig{file=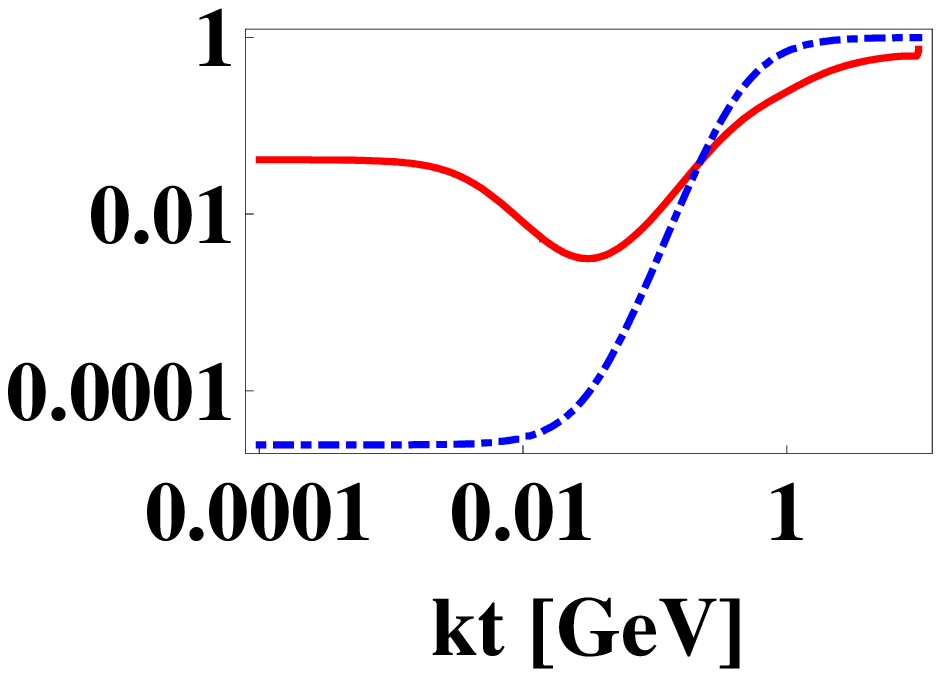,width=0.45\textwidth}
\end{center}
\caption{(Color online) The suppression of radiation (as compared to
  that from light quarks) at $x=0.1$ for charm quarks, left, and for
  bottom quarks, right. The full red line is the SQCD result,
  $\bar{K}$, and the dash-dotted blue line corresponds to the dead
  cone ratio $K_\mathrm{HQ}$, eq.~\ref{eq:KHQ}.}
\label{deadf3}
\end{figure}

In fig.~\ref{deadf3} we display this suppression of radiation from a
charm quark, left, and from a bottom quark, right, for $x=.1$ as a
function of $\kt$. The full red line is the SQCD result,
$\bar{K}(x=0.1,\kt;\mQ)$, and the dash-dotted blue line is the result
of the hard scattering approach, for which $\bar{K}=K_\mathrm{HQ}$,
see eq.~\ref{eq:KHQ}.  The suppression of the yield at small $\kt$ is
largely overestimated in the latter approach, but, more important for
the whole radiation, the SQCD ratio behaves as $\vkt^2$ at
intermediate $\kt$ when $x\mQ$ is comparable to $\mD$ (this is the
case for the bottom quark in fig.~\ref{deadf3}) instead of the
$(\vkt^2)^2$ rise of $K_\mathrm{HQ}$. For somewhat smaller $x\mQ$
(charm quark case in fig.~\ref{deadf3}), the rise after the dip is
comparable to that of $K_\mathrm{HQ}$ in the middle of the range but
there is an extra reduction visible for higher $\kt$. Looking at
fig.~\ref{deadf2}, the latter effect corresponds to the depletion of
the full curve relative to the dash-dotted curve. It is a consequence
of the mass effect in the region where $\kt$ is comparable to
$x\mQ$. Such a feature is already present before $\lt$ averaging where it
can be attributed to the departure from 1 of the last factor in
eq.~\ref{eq:soft_scatt_dist}.

These results show that the mass effect on radiation is more involved
than what can be modeled with a simple ``dead cone suppression''
factor. A similar conclusion was reached in Ref.~\cite{ASW}
in a situation that goes beyond one single averaged scattering.

\section{Radiation Cross Section and Power Spectrum}
\label{sec:rad_cross_section}

We now come back to the case of finite $s$ and perform successively the integration on $\vec{k}_t$ and $\vec{l}_t$ to obtain the
differential cross section $\frac{d\sigma^{Qq\rightarrow Qgq}}{dx}$
which describes the power spectrum.

\subsection{Integration over $\kt$}
\begin{figure}[H]
\begin{center}
\epsfig{file=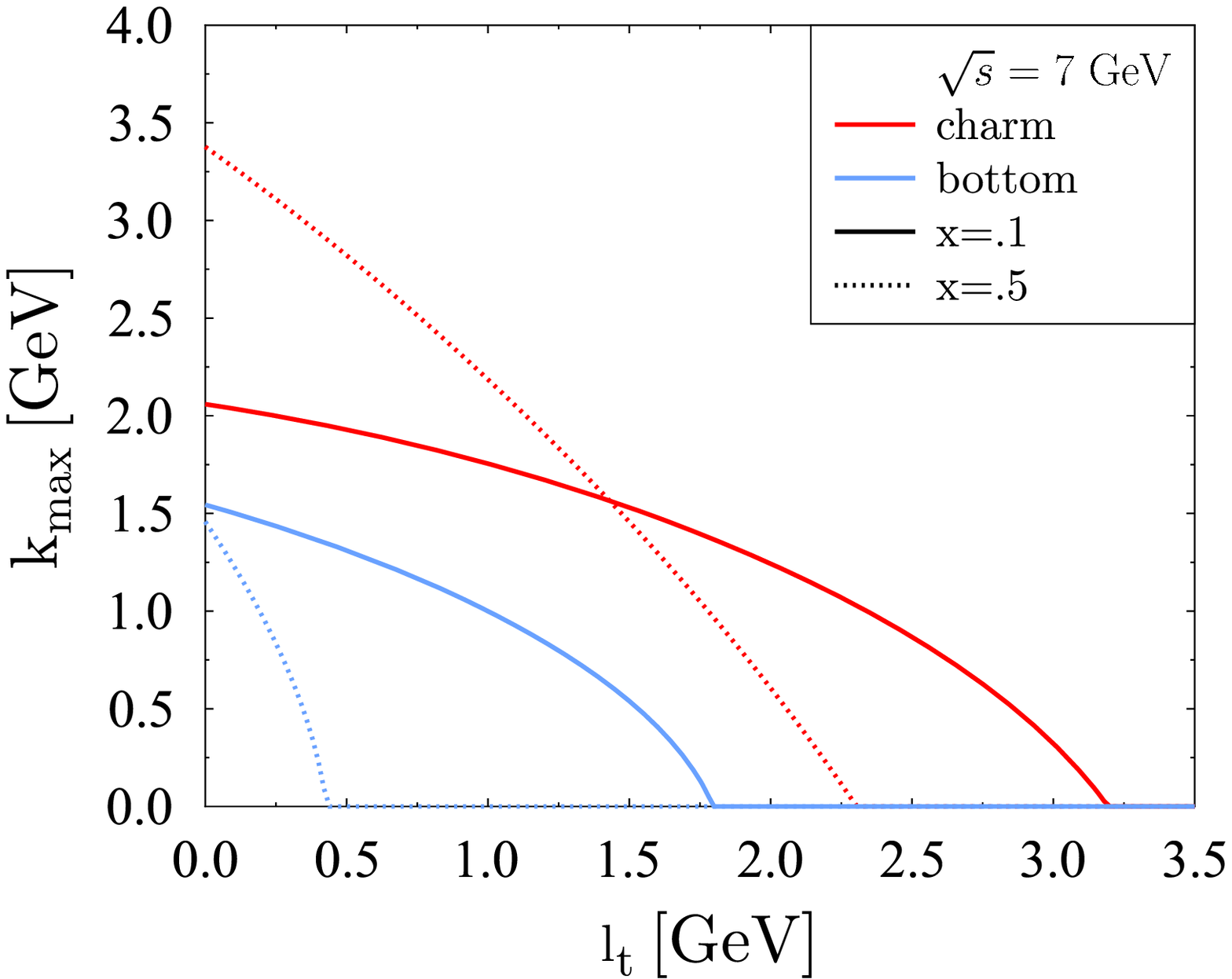,width=0.45\textwidth}
\epsfig{file=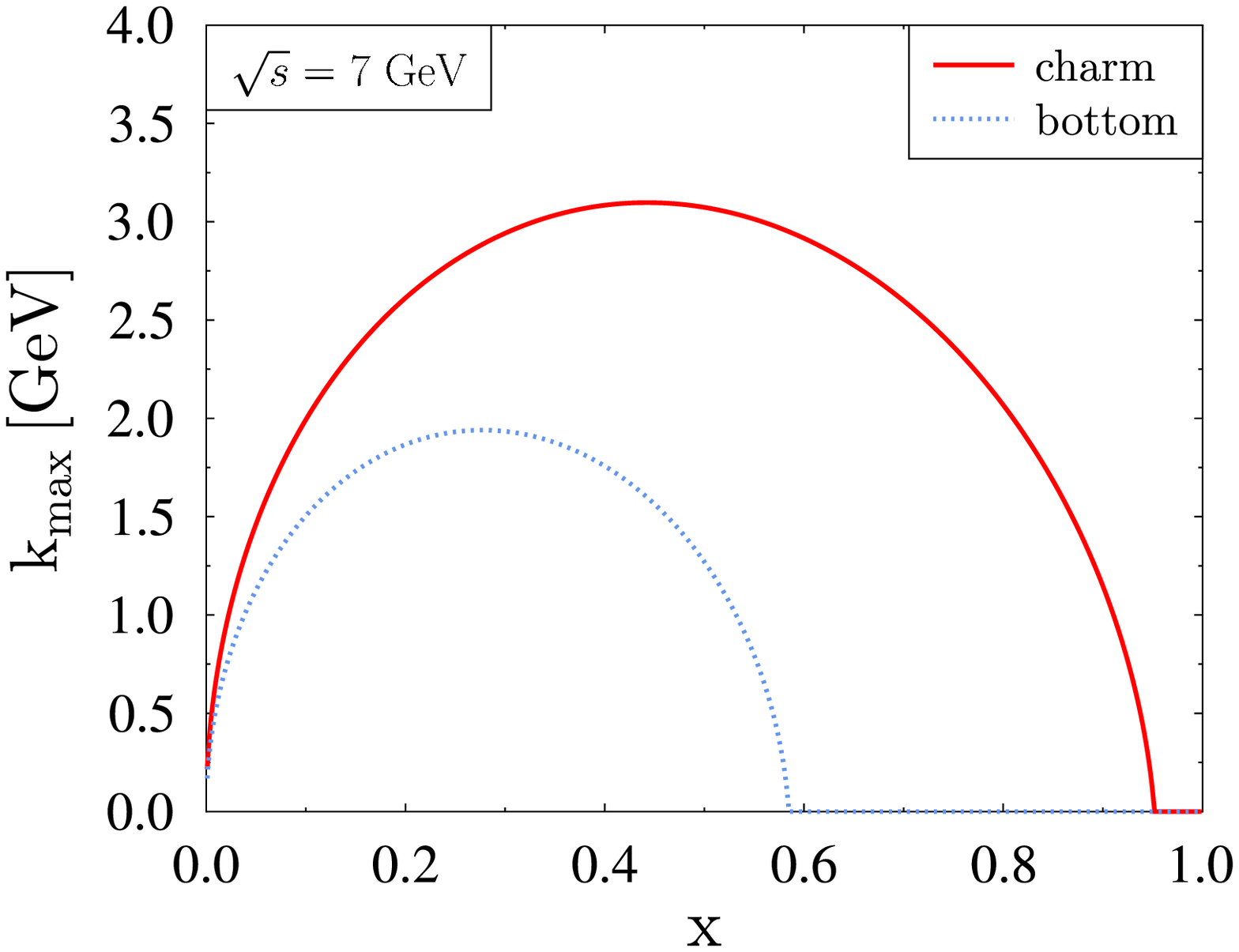,width=0.45\textwidth}
\end{center}
\caption{(Color online) The upper integration limit $k_\mathrm{max}$
  as a function of $\lt$ for $x=.1$ and $x=.6$ and for charm and
  bottom quarks (left) and as a function of $x$ for $\lt =.3$~GeV for
  charm and bottom quarks (right).}
\label{phpp}
\end{figure}

For the ``exact'' model, model I, performing the 
integration over $\kt$ is rather involved and a numerical approach
was preferred. In the case of model II, eq. \ref{def:model2}, the 
simple integrand makes it possible to perform the integration over $\kt$
analytically provided one neglects the angular dependence of
the phase space boundary, which is indeed rather mild under the conditions of
eq.~\ref{eq:hypsmalltransverse}.:
\begin{eqnarray}
\int\! P_g\,\Theta(K_t-\kt)  d^2\vkt
&=&\frac{C_A\alphas}{\pi}\frac{1-x}{x}
\left(-\frac{K_t^2}{K_t^2+x^2\mQsq}-\frac{1}{2}
-\frac{K_t^2-\vlt^2-x^2\mQsq}{
2\sqrt{K_t^4-2\left(\vlt^2-x^2\mQsq\right)K_t^2
+\left(\vlt^2+x^2\mQsq\right)^2}}\right.\nonumber\\
&+&\frac{\vlt^2+2x^2\mQsq}{\sqrt{\vlt^2(\vlt^2+4x^2\mQsq)}}
\left[\log\frac{\vlt^2+3x^2\mQsq+\sqrt{1+4x^2\mQsq/\vlt^2}\;
(\vlt^2+x^2\mQsq)}{x^2\mQsq}\right.
\label{ktinti}\\
&-&\left.\left.
\log\frac{-K_t^2+\vlt^2+3x^2\mQsq+\sqrt{1+4x^2\mQsq/\vlt^2}\,
\sqrt{K_t^4-2\left(\vlt^2-x^2\mQsq\right)K_t^2
+\left(\vlt^2+x^2\mQsq\right)^2}}{K_t^2+x^2\mQsq}\right]\right)\,,
\nonumber
\end{eqnarray}
Physics-wise, the expression for the upper limit of the $\kt$ integration $K_t\rightarrow k_{\rm max}(s,\lt)$ 
is given by the root of $\Delta$ in eq. \ref{eq:def_Delta}. It is shown in
fig.~\ref{phpp}, as a function of $\lt$ (left) and as a function of
$x$ (right). Except close to the boundaries, $x=0$ and
$x=x_\mathrm{max}=(s-\mQsq)/s$ the kinematically allowed values reach
several GeV for typical $\lt$, $\lt=\mathcal{O}(\mD)\le 1$~GeV. In
this situation, the high energy limit $K_t\to\infty$ should provide a
reasonable approximation of the full expression. Requiring $K_t\gg\mD$
at $x=x_\mathrm{max}/2$ translates into $s-\mQsq\gg 2\sqrt{s}\mD$. For
realistic numbers, we anticipate a failure of the high-energy
approximation in the whole $x$-range for $\sqrt{s}\le 2$~GeV,
respectively 5~GeV, in the case of a charm, respectively bottom,
quark.

In the $K_t\to\infty$ limit, eq.~\ref{ktinti} simplifies considerably:
\begin{equation}
\int P_g d^2\vkt  =\frac{C_A\alphas}{\pi}\frac{1-x}{x}
\left(\frac{\vlt^2+2 x^2\mQsq}{\sqrt{\vlt^2(\vlt^2+4x^2\mQsq)}}
\log\left(\frac{\vlt^2\left(\sqrt{1+4 x^2\mQsq/\vlt^2}+1\right)
+4 x^2\mQsq}{x^2\mQsq\left(
\sqrt{1+4 x^2\mQsq/\vlt^2}-1\right)}+1\right)-2\right).
\label{klimno}
\end{equation}

\begin{figure}[H]
\begin{center}
\epsfig{file=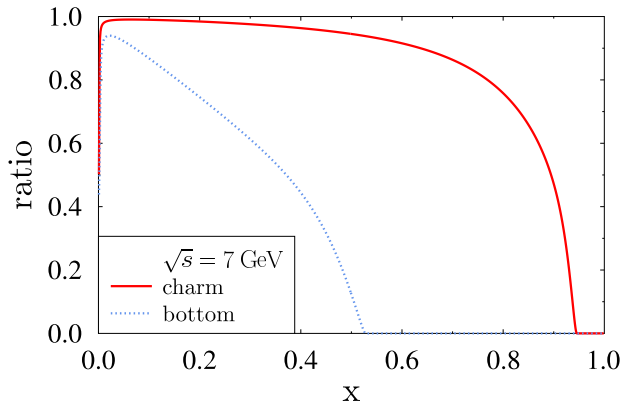,width=0.45\textwidth}
\end{center}
\caption{(Color online) Influence of the phase space limitation for
$\sqrt{s}=$7~GeV and $\lt=$ .3~GeV. We display the ratio
  $\int P_g \Theta(k_\mathrm{max}-\kt)d^2\vkt/\int P_g d^2\vkt$ , the right hand side of  
eq. \ref{ktinti} divided by the right hand side of eq. \ref{klimno}, for
  charm and bottom quarks.}
\label{ph1}
\end{figure}

In fig.~\ref{ph1} we show the ratio $\int P_g
\Theta(k_\mathrm{max}-\kt)d^2\vkt/\int P_g d^2\vkt$ for
$\sqrt{s}=$7~GeV. The main effect is the absence of radiation in the
kinematically forbidden region $x\ge x_\mathrm{max}$, a characteristic
that is not present in the approximate expression
eq.~\ref{klimno}. Close to $x=0$ the phase space limit reduces the
integral but for finite values of $x$ up to the upper limit in $x$
both equations agree very well. We notice that the ratio at small $x$
should not be expected to go to 0 despite of a closure of phase space
such as $k^2_\mathrm{max}\sim x(s-\mQsq)$ since the effective lower
bound for large radiation goes to 0 even faster as $x^2\mQsq$. In
practice, this influence of the phase space boundary shows up when
$x^2 M^2\lesssim k_{\rm max}^2 \lesssim \vec{l}_t^2$. However, we
already mentioned that this very small $x$ region
$x=\mathcal{O}(\vec{l}_t^2/(s-\mQsq))$ is beyond the scope of the
present study.

The integrated gluon distribution, eq.~\ref{klimno}, depends on $\vlt^2$
and $x^2\mQsq$ through the ratio $\vlt^2/(x^2\mQsq)$ only. For hard
scattering, when this ratio is large, the limiting form of
eq.~\ref{klimno} is given by
\begin{equation}\label{eq:intPghard}
\int P_g d^2\vkt \sim 2\frac{C_A\alphas}{\pi}\frac{1-x}{x}
\log\frac{\vlt^2}{x^2\mQsq},
\end{equation}
which shows the logarithmic enhancement mentioned at the beginning of
section~\ref{sec:gluon_distribution}. For soft scattering, when the
ratio $\vlt^2/(x^2\mQsq)$ is small, the limiting form is
\begin{equation}\label{eq:intPgsoft}
\int P_g d^2\vkt \sim \frac{C_A\alphas}{\pi}\frac{1-x}{x}
\frac{2\,\vlt^2}{3\,x^2\mQsq}.
\end{equation}
The proportionality of the result to $\vlt^2/(x^2\mQsq)$ is evident
from the approximate constancy of $\mathcal{I}$ in this regime,
eq.~\ref{eq:soft_scatt_dist}, up to $\vkt^2\sim x^2\mQsq$. The hard
scattering approximation of the spectrum,
eq.~\ref{eq:hard_scatt_dist}, supplemented with a cut-off
$\vkt^2<\vlt^2$, is completely off as it would result in a
proportionality of the result to $(\vlt^2/(x^2\mQsq))^2$. This sheds a
complementary light to the discussion carried out in
section~\ref{sub:small_kt} on the dead cone effect. In
eq.~\ref{eq:intPgsoft} the radiation is in proportion to the
square of the transverse momentum transfer, as expected. 
It is comparatively weak as compared to the hard scattering regime as a
consequence of a strong destructive interference in the soft regime. A
simple interpolation between these two limiting forms has been
advanced in~\cite{Peigne:2008wu}
\begin{equation}
\int P_g d^2\vkt \approx 2\frac{C_A\alphas}{\pi}\frac{1-x}{x}
\log\left(1+\frac{\vlt^2}{3\,x^2\mQsq}\right).
\end{equation}
This expression approximates the full result, eq.~\ref{klimno}, with a
deviation smaller than 3\% over the full range of
$\vlt^2/(x^2\mQsq)$. This is the approximation we will consider in all
subsequent comparisons.

\subsection{Power spectrum}
\label{sub:wloss}

In order to calculate the power spectrum we come back to the
radiation cross section eq.~\ref{eq:xsec}. We have to integrate it
first over $\vkt$ as detailed above for the gluon distribution $P_g$
and next over the momentum transfer $\vlt$. Thus, from
\begin{equation}
\frac{d\sigma}{dx}=\int \frac{d\sigma^{Qq\to Qgq}}{dxd^2\kt d^2\lt}
d^2\vkt d^2\vlt,
\end{equation}
we obtain the fractional momentum loss spectrum $x\,d\sigma/dx$. At
high-energy, in the frame of the heat bath where non zero components
of the target parton momentum $q$ are of order $T$, we have
\begin{equation}
P^0\approx p^0\quad\mathrm{and}\quad k^0\approx x\,p^0,
\end{equation}
(even at small $x\sim k_q\equiv \vkt^2/(2x p\cdot\pb)$, we benefit from
the strong hierarchy $p^0\gg q^0$ in this frame) thus
\begin{equation}
x\frac{d\sigma}{dx}\approx
\left(\frac{k^0}{P^0}\right)_{\mathrm{HB}}\frac{d\sigma}{dx},
\end{equation}
justifying the identification of $x\,d\sigma/dx$ as a (fractional) energy loss spectrum. 
For $k_\mathrm{max}\to\infty$ we obtain a simple formula for the fractional energy loss spectrum
\begin{equation}
\label{polelo}
x\frac{d\sigma}{dx}=\frac{4C_AC_F}{N_c}\alphas^3(1-x)
\frac{\log(\frac{3x^2\mQsq}{\mD^2})}{3x^2\mQsq-\mD^2}=
\sigma_{\rm el}\, \frac{dI}{dx}\,,
\end{equation}
where 
\begin{equation}
\sigma_{\rm el}=\frac{C_F}{2N_c}
\int\frac{4\alphas^2}{(\vlt^2+\mD^2)^2}\;d^2\vlt
=\frac{4\pi C_F\alphas^2}{2N_c\mD^2}
\end{equation}
is the elastic $Qq\to Qq$ cross section and where
\begin{equation}
\frac{dI}{dx}=\frac{2C_A}{\pi}\alphas(1-x)
\frac{\log(\frac{3x^2\mQsq}{\mD^2})}{\frac{3x^2\mQsq}{\mD^2}-1}
\end{equation}
is the differential fractional energy loss spectrum per elastic collision.
At small $x$ the hard scattering regime may be recognized with a
behavior that can be traced back to that of the integrated gluon
distribution, see eq.~(\ref{eq:intPghard}). At larger $x$ the soft
scattering regime takes over with a power-law suppression $\sim
1/x^2$, while the factor $1-x$ comes as an additional suppression factor. 
The transition between the two regimes is at
$x_\mQ=\frac{1}{\sqrt{3}}\mD/\mQ$.

The cross section then allows for calculating an
approximate value of the energy loss per unit length due to radiation, 
assuming all partons of the medium to be quasi static (and neglecting coherence effects of the 
LPM type):
\begin{equation}
\frac{dE_{\rm rad}}{dz}\approx \rho \int \omega \frac{d\sigma}{d\omega} 
d\omega \approx \rho E_{\rm beam} \int x \frac{d\sigma}{dx} dx.
\label{dedz}
\end{equation} 

The integral $\int x\frac{d\sigma}{dx} dx$ has no simple form but
behaves as $\sim (\mD/\mQ)\times 1/\mD^2$. This scaling law can be worked
out by breaking the integral of $\frac{dI}{dx}$ into two pieces, the
first encompassing the hard regime and the second for the soft
one. Making the appropriate approximations in both regimes, it is
found that both parts contribute equally when $\mD\ll\mQ$, leading
quantitatively to $\int \frac{dI}{dx} dx\approx
4\,x_\mQ\,(2C_A\alphas/\pi)$ up to a factor increasing from 0.5 when
$\mQ=\mD$ to 1.2 when $\mQ\to\infty$. For a small quark mass
$\mQ<\mD$, i.e. for larger values of $x_\mQ$, the dependence becomes
logarithmic. From this discussion it is clear that the radiation
depends strongly on the infrared cut-off $\mD$ introduced in the
matrix element for the elastic collisions.

\begin{figure}[H]
\begin{center}
\epsfig{file=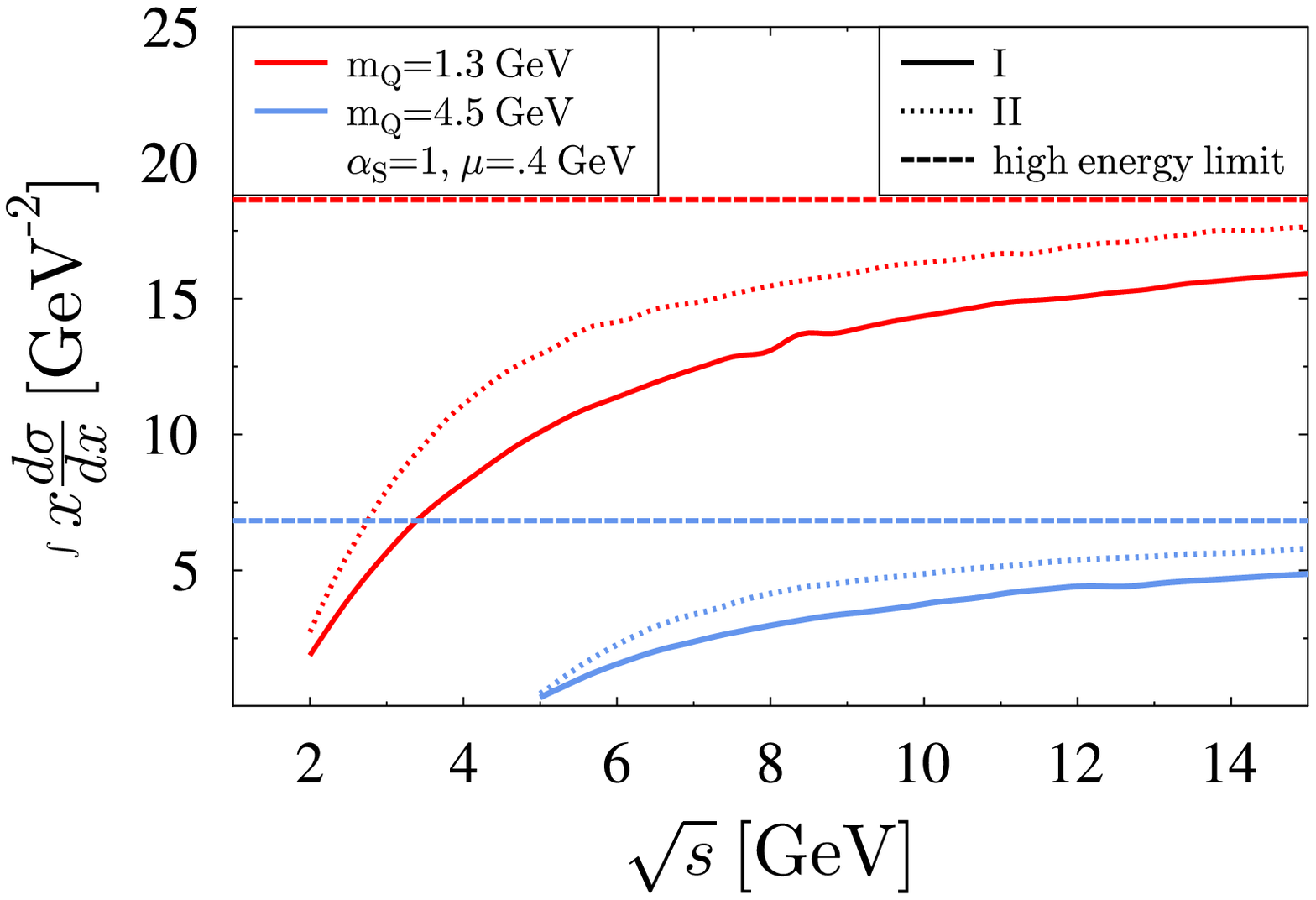,width=0.45\textwidth}
\epsfig{file=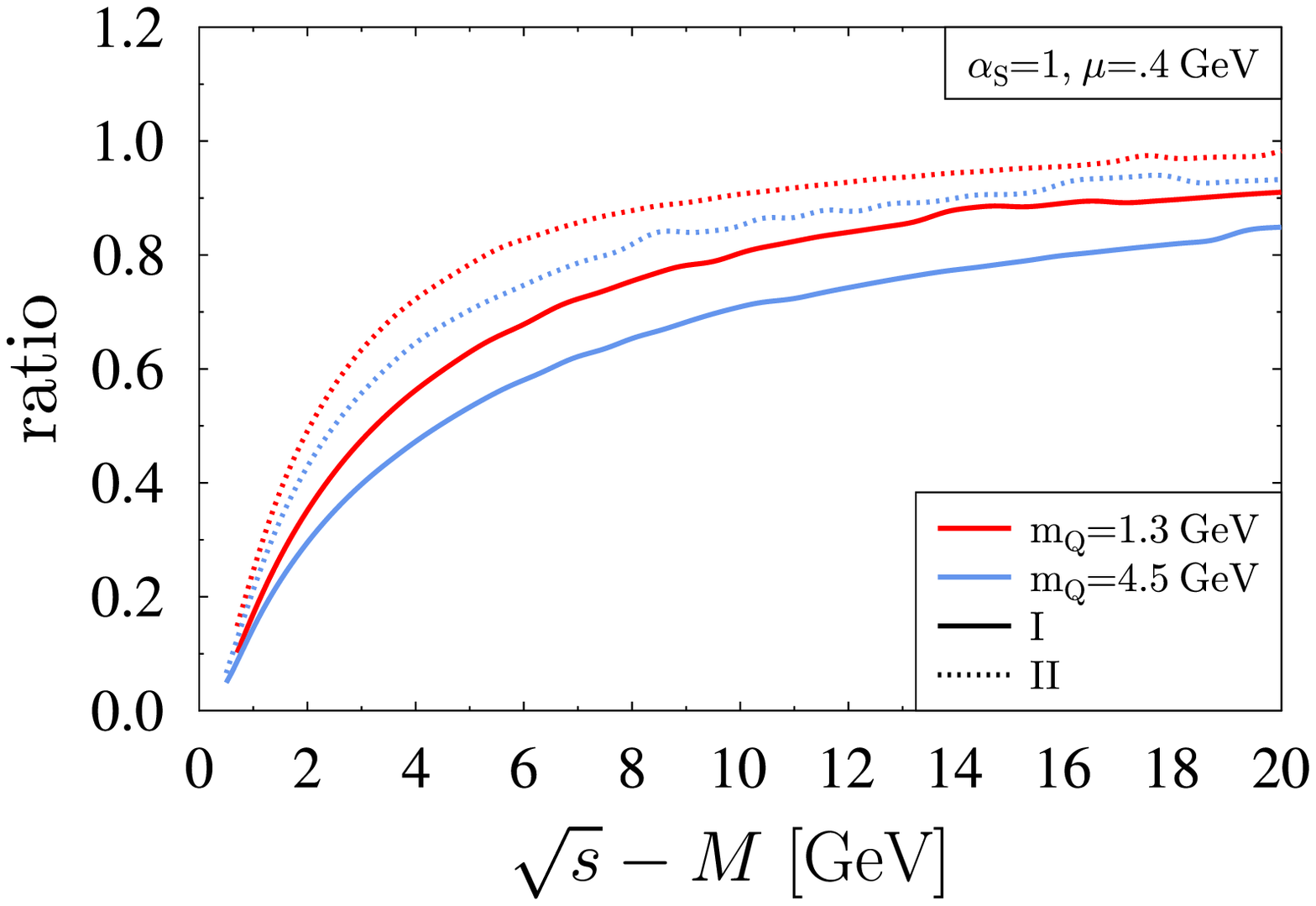,width=0.45\textwidth}
\end{center}
\caption{(Color online) Left: Integrated fractional energy loss $\int x
  \frac{d\sigma}{dx}\,dx$ as a function of $\sqrt{s}$ for models~I and~II
as well as for the high energy limit eq. \ref{polelo}.
  Right: The same divided by the high-energy limit as a function
  of $\sqrt{s}-\mQ$.}
\label{fig:intxdsigdx}
\end{figure}

In fig.~\ref{fig:intxdsigdx} left, the integrated fractional energy
loss $\int (x d\sigma/dx)\,dx$, corresponding to the energy loss per
unit length normalized to $\rho E_{\rm beam}$, is shown as a function of
$\sqrt{s}$. Model~I and model~II take finite $s$ 
corrections into account (see Sect.~\ref{sub:diff_x_sec})
and are compared to the $s$-independent integral of the high-energy approximation eq.~(\ref{polelo}). 
For the charm quark the full calculation, model~I,
reaches 50\% of the high-energy limit at $\sqrt{s}=4.5$~GeV and 75\%
at $\sqrt{s}=10$~GeV. It is also seen that model~II sits halfway
between model~I and the asymptotic result. In the right panel, the
same quantity normalized to the high-energy limit is displayed as a
function of $\sqrt{s}-\mQ$ in order to better judge the influence of
the heavy quark mass. As it turns out, the phase space limitation explains roughly half 
of the difference between model I and the high-energy limit and has the advantage to allow for an 
easy implementation.

\begin{figure}[H]
\begin{center}
\epsfig{file=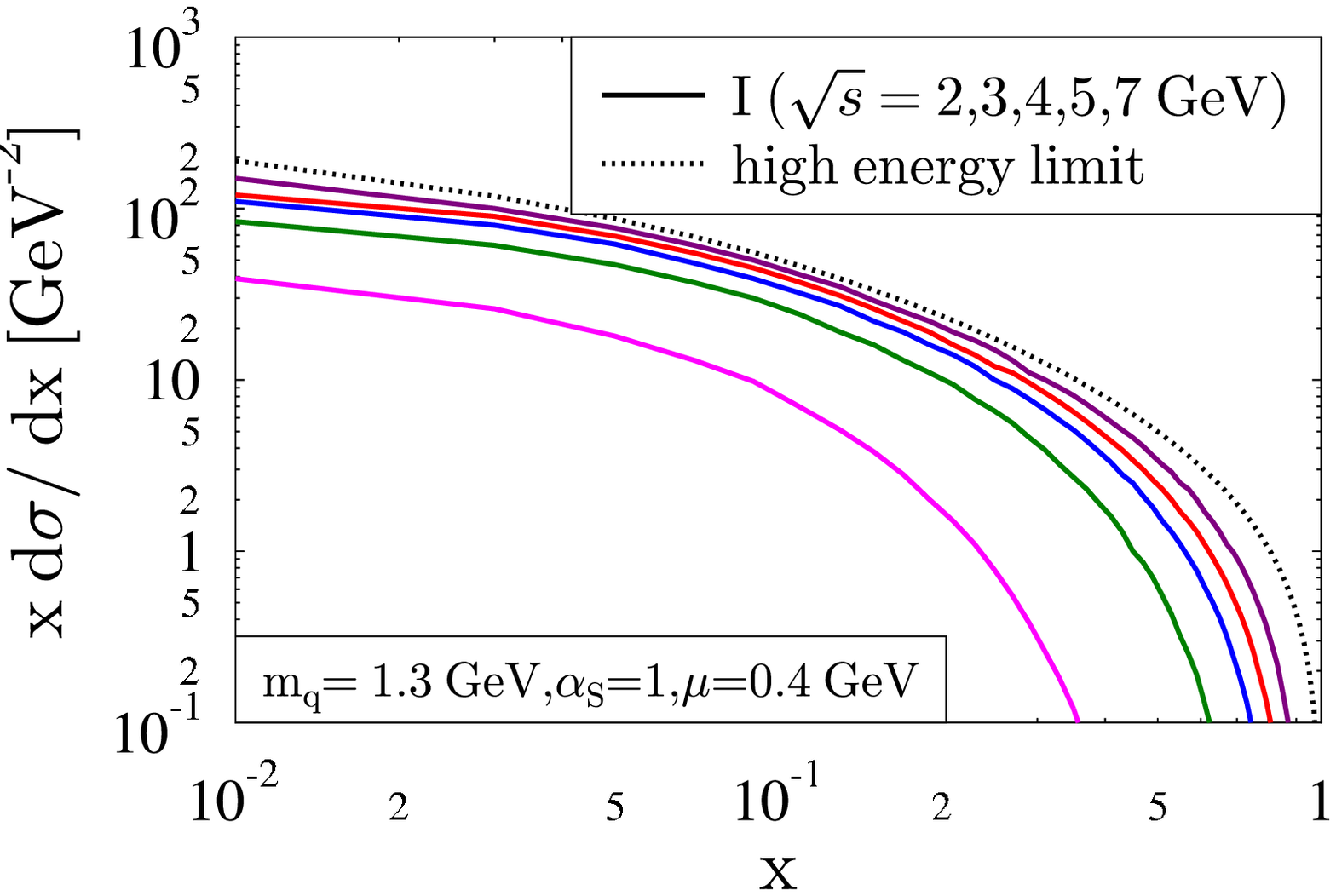,width=0.45\textwidth}
\epsfig{file=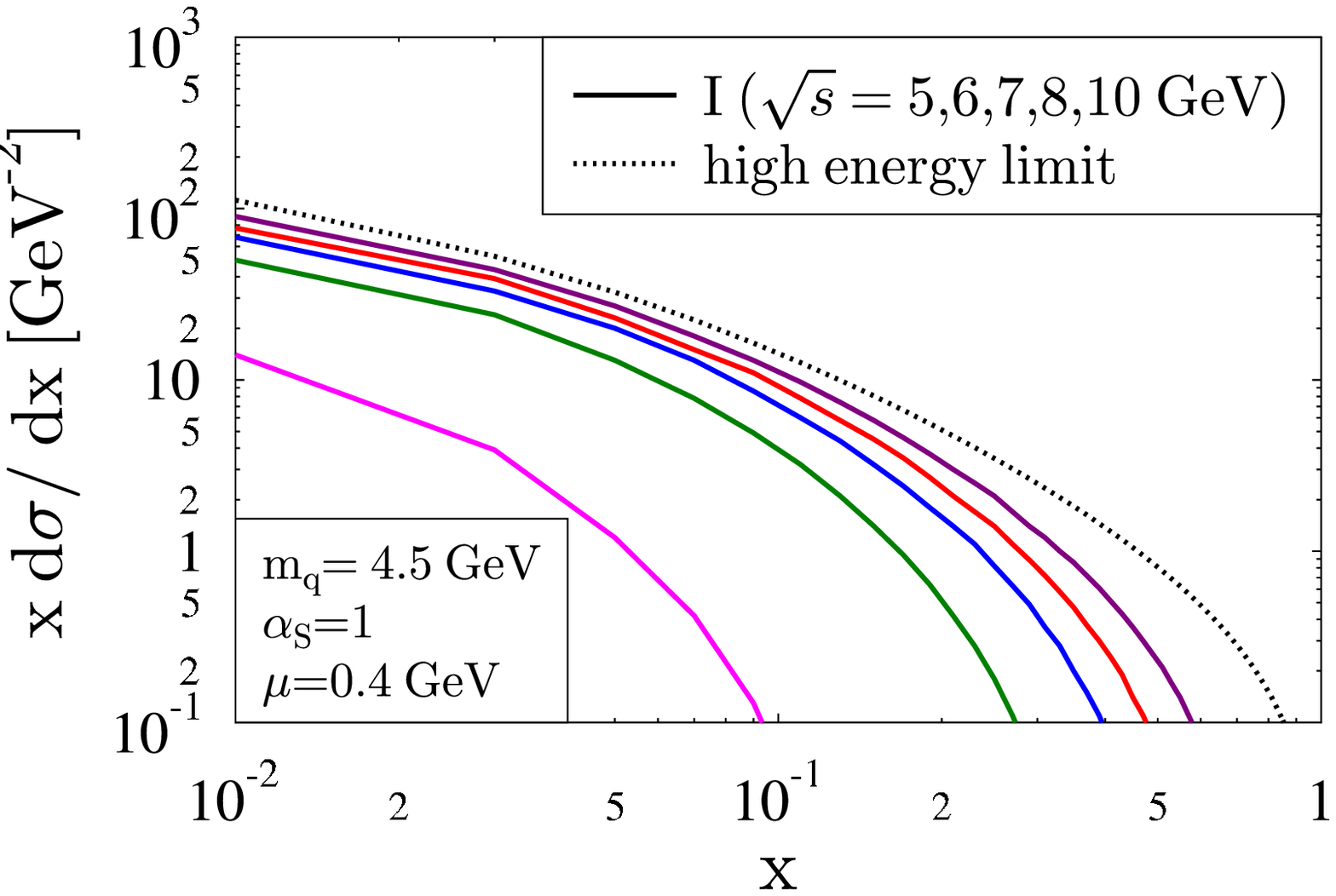,width=0.45\textwidth}
\end{center}
\caption{(Color online) The gluon emission cross section,
  $x\frac{d\sigma}{dx}$, for a charm quark (left) and a bottom quark
  (right) for several values of $\sqrt{s}$. We display the results of model~I and of the high-energy
  formula eq.~\ref{polelo}.}
\label{fig:xdsigdx}
\end{figure}
Fig.~\ref{fig:xdsigdx} shows the $x$-weighted differential energy loss
cross section $xd\sigma/dx$ calculated with the full SQCD matrix
element squared as compared to the approximation eq.~\ref{polelo}. 
The calculation is performed for several values of
$\sqrt{s}$. We observe that the approximate spectrum, eq.~\ref{polelo},
describes the trend, shown by the SQCD calculation, quite well at low
$x$ values, $0.01\le x\le 0.2$, where the deviation is roughly
constant and less than 50\% for $\sqrt{s}$ above 4.5~GeV. This explains the
result seen in fig.~\ref{fig:intxdsigdx}, remembering that the median
of $xd\sigma/dx$ is roughly given by $x_\mQ$ (about $0.2$ for charm
and $0.05$ for bottom). The phase space suppression at large $x$ plays an even larger role
and makes the emission of an energetic gluon an even
rarer process for the complete spectrum than it is when finite-$s$ corrections are
ignored.

\begin{figure}[H]
\begin{center}
\epsfig{file=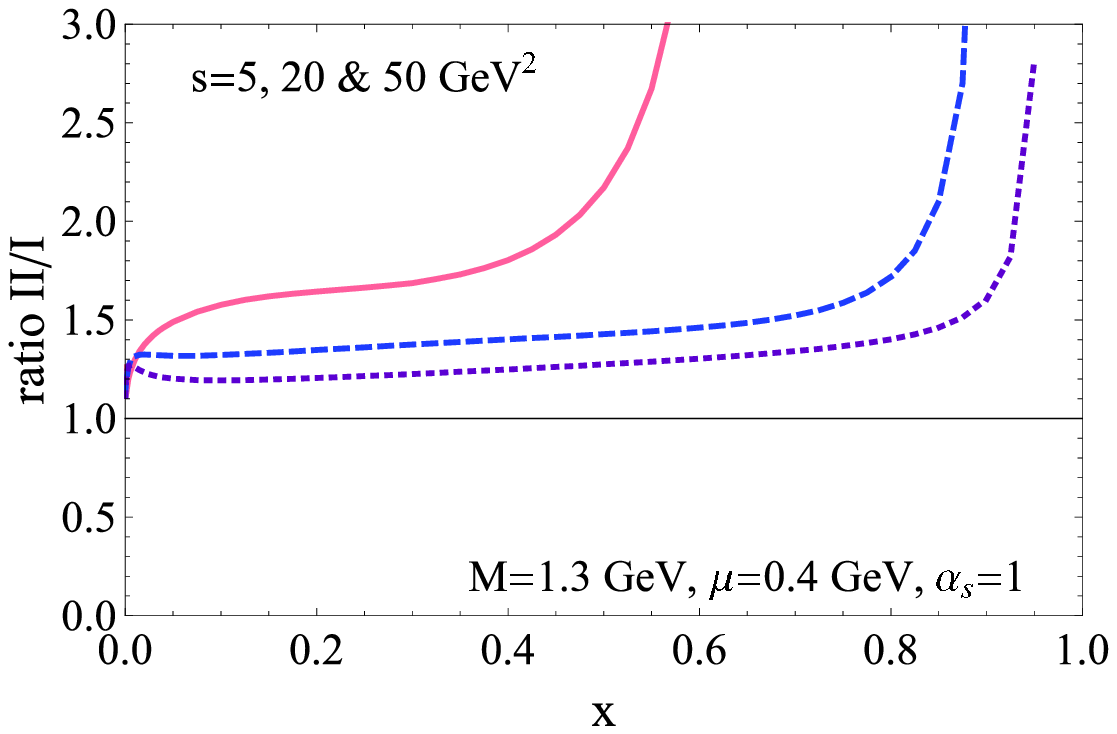,width=0.45\textwidth}
\epsfig{file=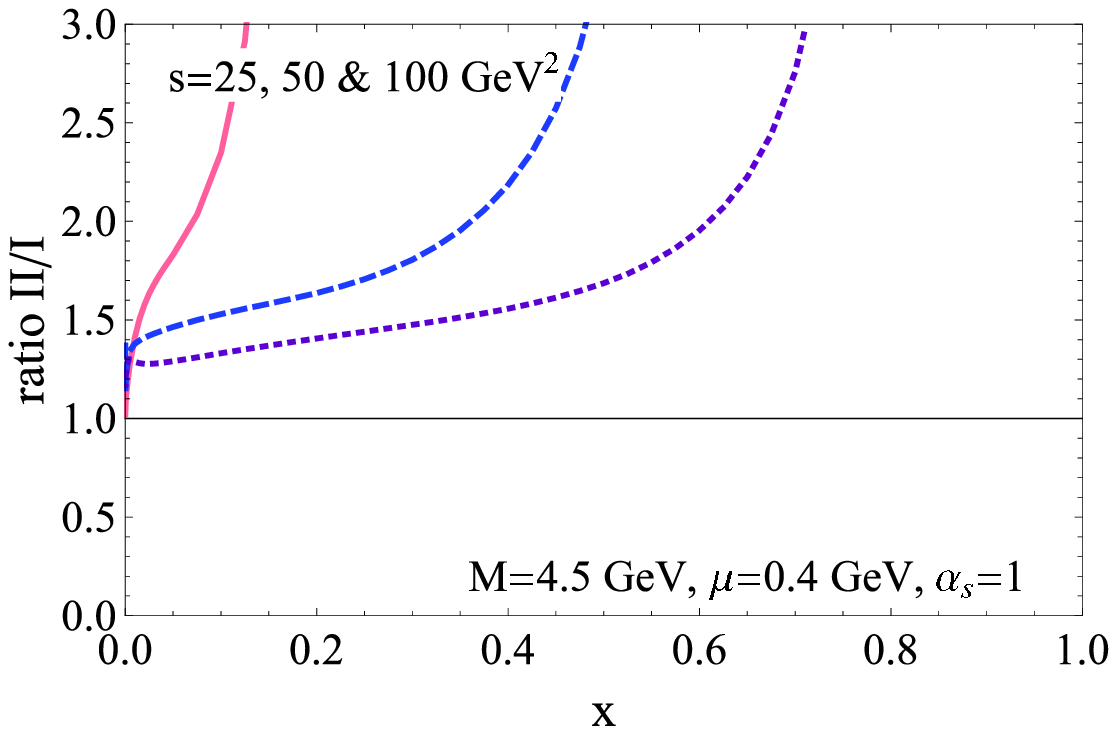,width=0.45\textwidth}
\end{center}
\caption{(Color online) Ratio of the result of model~II to the result of model~I for the quantity
  $x\frac{d\sigma}{dx}$ for charm (left) and bottom (right), plain and short-dashed lines
  correspond to the smallest and the largest values of $s$ in the set,  respectively.}
\label{fig:xdsigdx_ratio}
\end{figure}

The semi-quantitative description of the full result by the
high-energy approximation is welcomed for the phenomenology of
quenching which is dominated by the emission of low energy
gluons~\cite{Baier:2001yt}. Fig.~\ref{fig:xdsigdx_ratio} shows the ratio
model~II over model~I, for the quantity $x\frac{d\sigma}{dx}$, as a
function of $x$ and for several center of mass energies.
We observe that model~II (in which only the boundaries of the $(x,\vkt,\vlt)$ space are $s$-dependent) 
provides a reasonable approximation of the full
result, model~I, in a large interval of $x$, except at large $x$ or at
rather small energies.

\section{Finite gluon mass}
\label{sec:gluon_mass}

In the plasma the gluon is not a free particle since it is in
interaction with the plasma environment. Lacking a tractable theory of
how this modifies the rules used to compute the collisional and the
radiative cross sections we are bound to speculate on the main
phenomenological effects that could modify our results so far. In
Ref.~\cite{Gossiaux:2008jv}, for the study of collisional losses
within an approach inspired by that of Braaten and
Thoma~\cite{Braaten:1991jj}, the effect of Debye screening was found
to have a large impact on the values obtained for the energy loss. In
the above results, the regularization of the elastic amplitude with
the introduction of a mass parameter $\mD$ mimicked in a certain way
the screening phenomenon due to a gluon thermal mass. Of course, the
consequences of the occurrence of a thermal mass is not limited to the
addition of a regulator in some of the propagators. With the aim of
making a comparison between collisional and radiative energy losses
where thermal mass effects are treated on a similar footing we want to
explore the possible importance of a gluon mass on radiation.

To do so we employ a simple approach. We start out from the SQCD
matrix elements and retain only the dominant terms of the series
expansion in $\sqrt{s}$. Then we assume that the gluons have a finite
mass $k_\mu k^\mu= m_g^2$. Going through this calculation we find that
a finite gluon mass modifies only the denominator of the SQCD matrix
elements and we have to replace in eqs.~\ref{eq:M2SQCD}
and~\ref{eq:M2SQED}
\begin{equation}
x^2\mQsq\to x^2\mQsq+ (1-x)m_g^2.
\end{equation}
The modification of the phase space can by found in
Appendix~\ref{app:phase_space}. These modifications provide an
extension of model~II for finite $m_g$. Starting out from the hard
thermal loop propagators Djordjevic et al.~\cite{Djordjevic:2007at}
arrived recently at a similar conclusion.

\begin{figure}[H]
\begin{center}
\epsfig{file=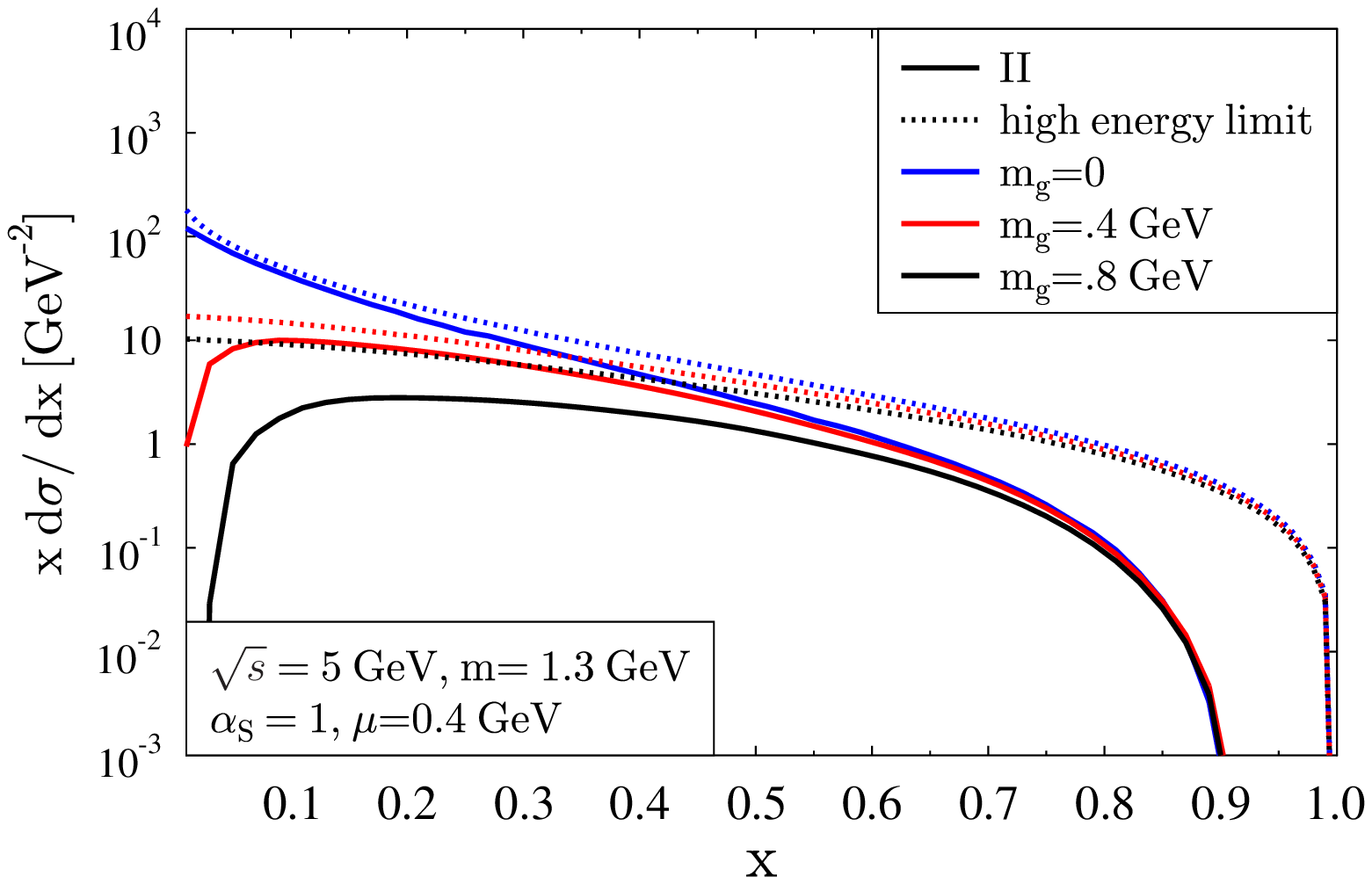,width=0.45\textwidth}
\epsfig{file=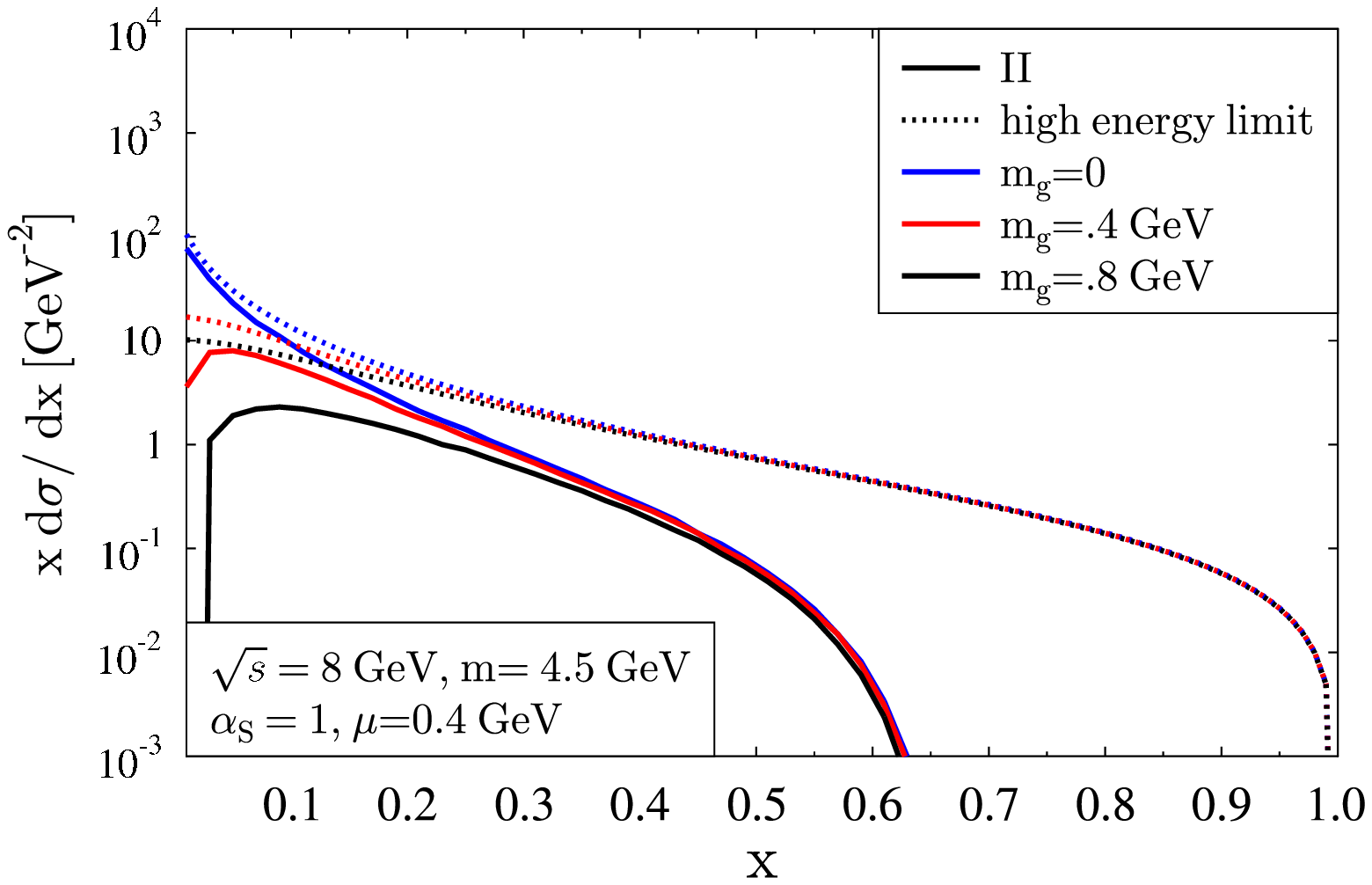,width=0.45\textwidth}
\end{center}
\caption{(Color online) Comparison of the approximate formula
  (eq.~\ref{polelomg}) as compared to the  SQCD solution for high
  energies and for different gluon masses $m_g$. Left: $xd\sigma /dx$
  of charm quarks at $\sqrt{s}=5$~GeV, right:  $xd\sigma /dx$ of 
  bottom quarks at $\sqrt{s}=8$~GeV.}
\label{xdpdxmgcomp}
\end{figure}

Using a finite gluon mass the matrix element for gluon emission,
eq.~\ref{eq:M2SQCD}, does not diverge for $\kt\to 0$ or $\kt\to\lt$,
even for the massless case $M=0$. Moreover, the energy loss at small
$x\lesssim m_g/M$ -- that is to say in the region of hard
scattering and dead cone effect -- is strongly reduced. In addition, finite gluon
masses reduce the phase space and very small values of $x$ are
kinematically not allowed anymore. This is seen in
fig.~\ref{xdpdxmgcomp} which shows for a gluon mass of $m_g= 0, 0.4$
and $0.8$~GeV the $x$-weighted differential gluon cross section
$x\frac{d\sigma}{dx}$. The left panel shows the cross section for
charm quarks at $\sqrt{s}=5$~GeV, the right panel that for bottom
quarks at $\sqrt{s}=8$~GeV. One can extend the approximative formula,
eq.~\ref{polelo}, also toward finite gluon masses
\begin{equation}
x\frac{d\sigma}{dx}=4C_F\alphas^3(1-x)
\frac{\log(\frac{3(m_g^2+x^2\mQsq)}{\mD^2})}{3(m_g^2+x^2M^2)-\mD^2}.
\label{polelomg}
\end{equation}
This formula is compared in fig.~\ref{xdpdxmgcomp} with the SQCD
calculation in the high energy limit. Good agreement is found for a
small gluon mass only. The phase space limitations at small $x$,
caused by a finite gluon mass, are indeed not contained in
eq. \ref{polelomg}.

\begin{figure}[H]
\begin{center}
\epsfig{file=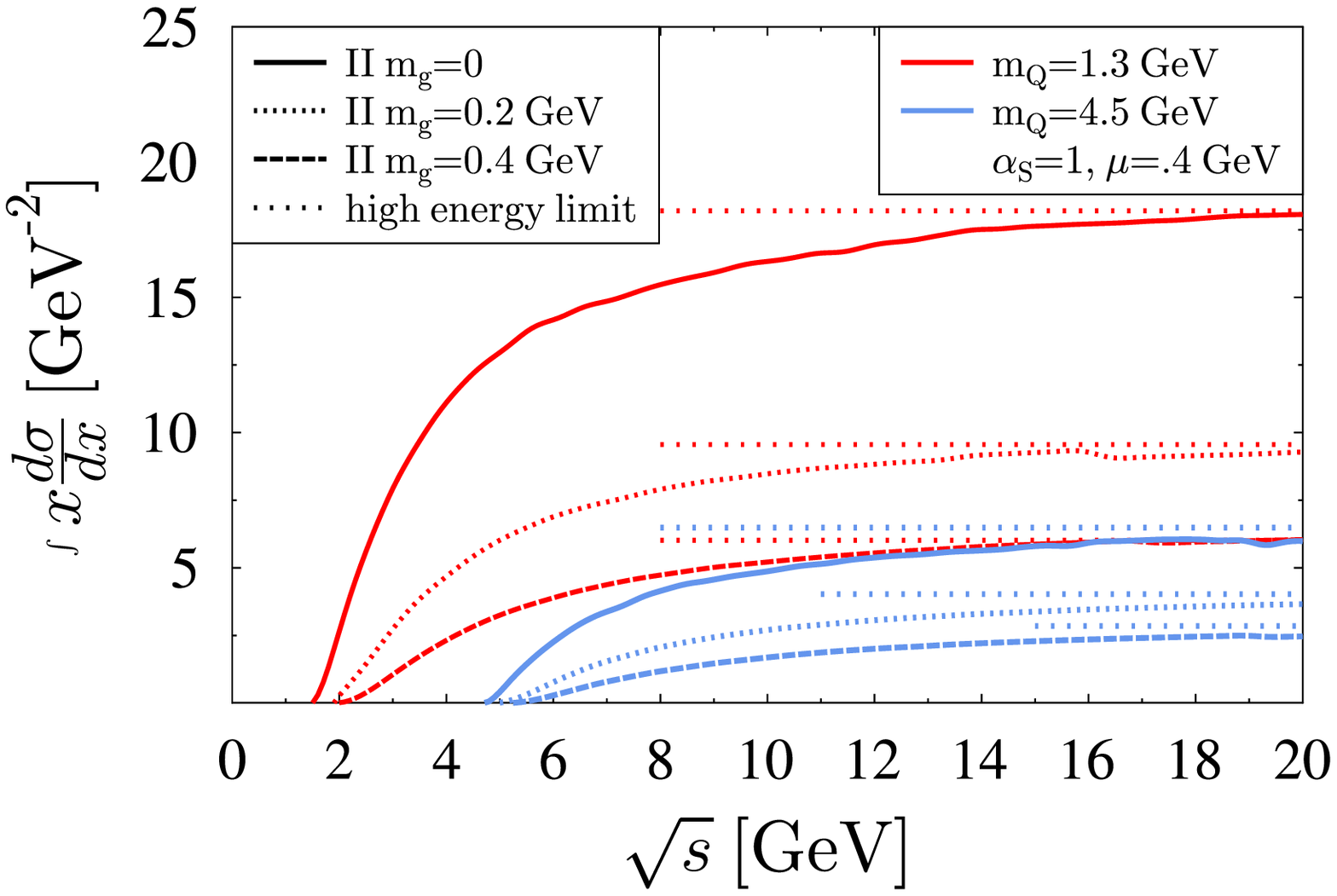,width=0.45\textwidth}
\epsfig{file=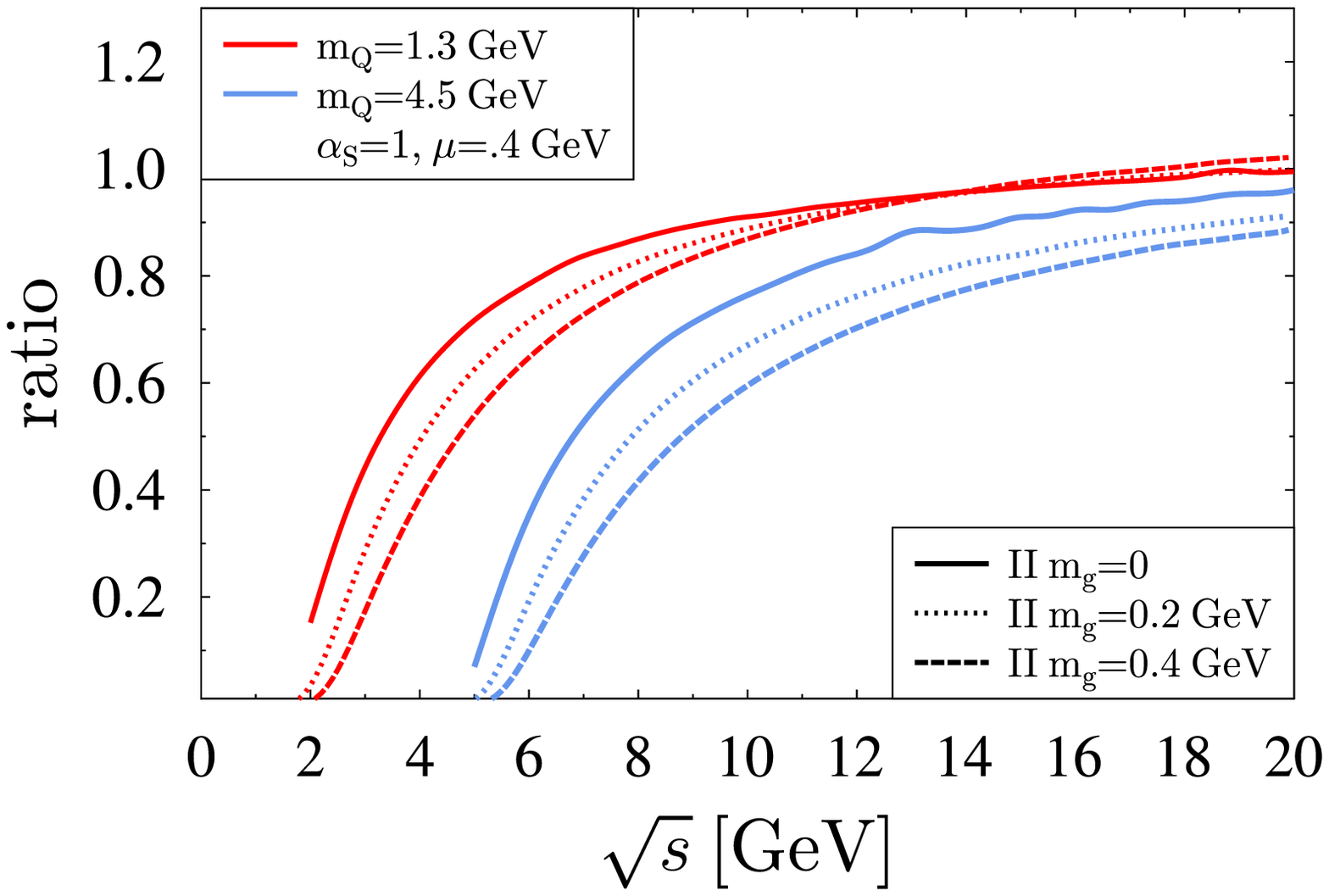,width=0.45\textwidth}
\end{center}
\caption{(Color online) Left: integrated fractional energy loss $\int x
  d\sigma/dx\,dx$ as a function of $\sqrt{s}$ for model~II with
  $m_g=0, 0.2,0.4$~GeV and $\mQ=1.3,4.5$~GeV. Right: model~II divided
  by the high-energy limit as a function of $\sqrt{s}$.}
\label{fig:intxdsigdx_mgneq0}
\end{figure}

The effect of a finite $m_g$ on the integrated fractional energy loss
$\int (x d\sigma/dx)\,dx$ is shown in fig.~\ref{fig:intxdsigdx_mgneq0}.
Although the gluon mass has a large effect on the absolute values, 
it affects only mildly the normalized quantities obtained by dividing 
by the high-energy limit.

\section{Energy loss}
\label{sec:energy_loss}

To close this investigation we compare the radiative and collisional
energy loss of heavy quarks induced by scattering on light ones. 
To evaluate those, we start from the covariant expression for the infinitesimal 
evolution of the average 4-momentum:
\begin{equation}
\langle \frac{-dP^\mu}{d\tau}\rangle=\frac{1}{M}\int \frac{d^3q}{q^0}
P\cdot q f(q) \int \frac{d\sigma}{dQ}(P-P')^\mu dQ,
\label{exact_eloss}
\end{equation}
where $\tau$ is the heavy quark proper time, $dQ$ is the invariant phase space corresponding to the exit channel and $f$ is the invariant Fermi-Dirac distribution of light quarks 
\begin{equation}
f(q) = \frac{g_q}{e^{(q\cdot u_{\rm HB} - \mu_q)/T}+1},
\end{equation}
with $g_q$ the degeneracy factor and $u_{\rm HB}$ the heat bath 4-velocity. To evaluate the average energy loss in the heat bath frame, we project eq.~\ref{exact_eloss} on $u_{{\rm HB},\mu}$ and then proceed to the calculation of the integrals in the rest frame (r.f.) of the heavy quark.
Concentrating on the energy lost through gluon radiation, eq.~\ref{exact_eloss} simplifies to
\begin{equation}
\langle -\frac{dE_{\rm rad}}{d\tau}\rangle=\frac{M}{2}\int 
d^3q f_{\rm r.f.}(\vec{q}) \int \frac{d\sigma}{dQ}
\left[(u^0_{\rm HB}+|\vec{u}_{\rm HB}| 
\cos\theta{(\vec{u}_{\rm HB},\vec{q})}) x
+ (u^0_{\rm HB}-|\vec{u}_{\rm HB}| \cos\theta{(\vec{u}_{\rm HB},\vec{q})}) 
\frac{\vec{k}_t^2}{x M^2}\right]dQ  
\label{exact_eloss_simpl}\,,
\end{equation}
where $u_{\rm HB}$ is here the heat bath velocity measured in the HQ rest frame, $u_{\rm HB}=(P^0,-\vec{P})/M$. 
As $f_{\rm r.f.}(\vec{q})$ is centered around $\vec{q}$ such that 
$(u^0_{\rm HB}-|\vec{u}_{\rm HB}| \cos\theta{(\vec{u}_{\rm HB},\vec{q})})=0$, while 
$\int \frac{\vec{k}_t^2}{x M^2} \frac{d\sigma}{dQ} dQ$ is generically smaller
than $\int x \frac{d\sigma}{dQ} dQ$ for $x\in[\mu/M,1]$ which represents
the largest fraction of the integration domain, we neglect the second term in the bracket of eq. \ref{exact_eloss_simpl}. 
Using $d\tau=\frac{M}{P^0} dt$, we then arrive at
\begin{equation}
\langle -\frac{dE_{\rm rad}}{dt}\rangle\approx \frac{M}{2}\int 
d^3q f_{\rm r.f.}(\vec{q}) \left(1+\frac{|\vec{P}|}{P^0} \cos\theta{(\vec{u}_{\rm HB},\vec{q})}\right) \int  x \frac{d\sigma}{dx} dx\,,
\end{equation}
where the last factor has been discussed in the previous sections.
It is then trivial to perform the angular integration on 
$\cos\theta{(\vec{u}_{\rm HB},\vec{q})}$ and to express the energy loss as a simple convolution on the $s$ variable. We proceed similarly for the collisional
energy loss, using the SQCD expression for $\frac{d\sigma_{\rm el}}{dQ}$.

\begin{figure}
\begin{center}
\epsfig{file=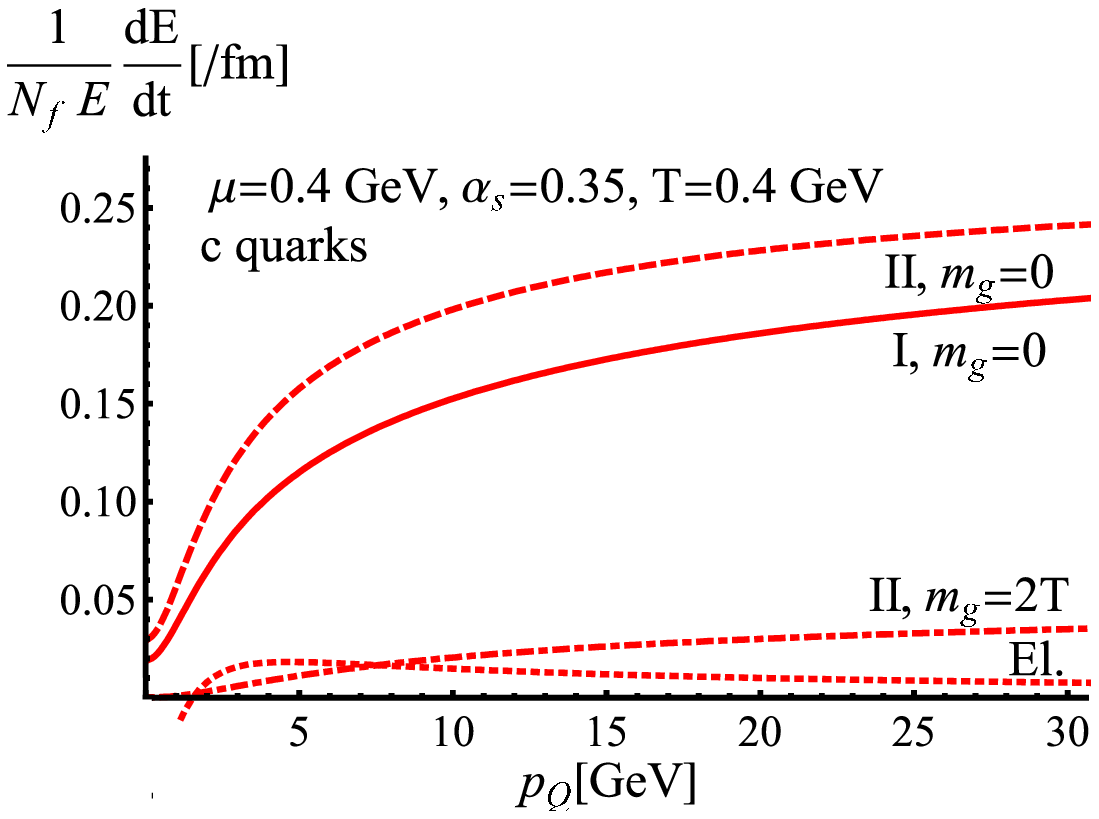,width=0.45\textwidth}
\epsfig{file=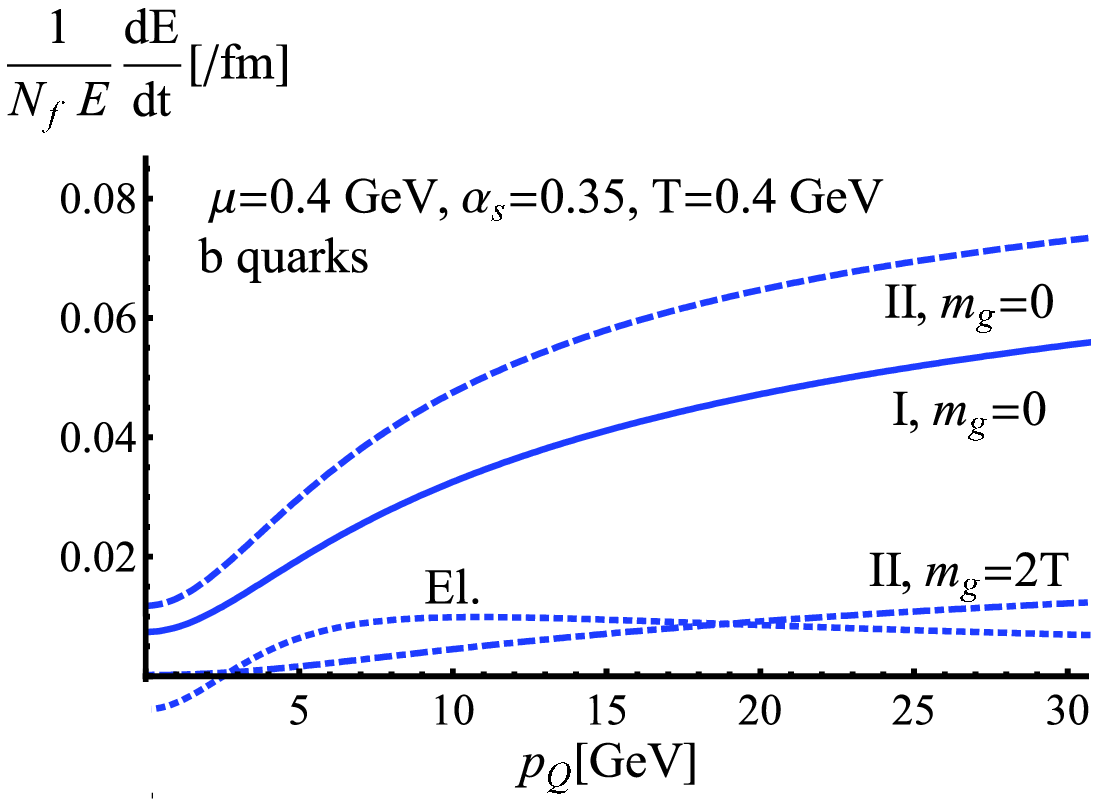,width=0.45\textwidth}
\end{center}
\caption{(Color online) Comparison between radiative and collisional energy loss 
of charm quarks (left) and of bottom quarks (right). Only the contribution from collisions with the light quarks of the heat bath is shown.}
\label{peismi}
\end{figure}

In fig.~\ref{peismi}, we illustrate $dE/dt$ per flavour degree of freedom normalized to the
heavy quark energy. A Fermi-Dirac distribution has been used, 
with a temperature $T=0.4~{\rm GeV}$ and 0 quark chemical potential $\mu_q$, while the screening mass $\mu$ was taken 
as $\sqrt{0.15}\,m_D\approx 0.4~{\rm GeV}$, in agreement with
the model C of \cite{Gossiaux:2008jv}. 
The plain and dashed lines show $m_g=0$ results with either model~I or~II 
for both charm and bottom quarks. Model~II with $m_g=m_D/\sqrt{2}\approx 2T=0.8$~GeV is 
displayed as dot-dashed line. The energy loss strongly depends on the
values of $\mD$ and $m_g$ and hence on the plasma environment in which
the heavy quark moves. For our choice $\mu^2=0.3 m_g^2$, the radiative 
energy loss for heavy quarks is even dominated by the
collisional energy loss (short-dashed lines) for $|\vec{P}|$ up to several times the
heavy quark mass.

The radiative energy loss is calculated for the interaction of the
heavy quark with the light quarks of the plasma. If one wants to add
the radiation due to the interaction af a heavy quark with gluons one
has to add the gluon density to the quark density in
eq.~\ref{dedz}. At high energy, this corresponds to a multiplication 
of the quark density by $(1+\frac{N_f}{6}) / \frac{N_f}{6}$ which stems from
the color factor and the flavor degrees of freedom (including quantum statistics) of $t$-channel 
elastic scattering of $Q-q$ and $Q-g$~\cite{Peigne:2008nd}.

\section{Conclusion}

We present in this paper an approach to describe gluon emission from a
heavy quark in collision with a light quark at mid and forward
rapidity, i.e. for $ x\gg k_t^2/s$.  Because this radiation is
centered at $x\sim k_t/\sqrt{s}$ and the correction are of ${\cal
O}(x^2)$ we use for this approach the scalar QCD formalism which
allows identifying the physical processes much easier.  We separate
the matrix elements into three gauge invariant subgroups where two of
them are identical with the bremsstrahlung diagrams already observed
in QED. In the paper we concentrate on the third group which is
genuine to QCD and dominates the radiation in the central rapidity region.

We compare the full result with a high energy approximation which can
be analytically calculated and find fair agreement already for
moderate $\sqrt{s}$ values ($\sqrt{s}-M\gtrsim 7 \ {\rm GeV}$).  In
ultrarelativistic heavy ion collisions at RHIC and LHC the typical
$\sqrt{s}$ value of collisions of heavy quarks with heat bath
particles is lower and therefore the approximate formulas are not
directly applicable. We show that the phase boundaries are responsible
for a substantial part of the corrections at intermediate energies, so
that a model based on asymptotic transition elements and exact phase
space boundaries can be used for semi-quantitative purposes.

We find that the mass of the heavy quark suppresses the emission of
gluons at low transverse momentum (dead cone effect) but this
suppression is less important and less universal than originally
advocated.  We study the influence of a finite gluon mass on the
energy loss of heavy quarks in radiative collisions and find a quite
strong dependence. For massless gluons, the energy loss of heavy
quarks due to radiative collisions exceeds that due to elastic
collisions for all heavy quark momenta, while for massive gluons the
crossing happens at moderate but finite momenta. In all cases, we
conclude that radiative collisions have to be included for a
quantitative description of the energy loss of heavy quarks in a quark
gluon plasma. We have carried out this study assuming  
a constant infrared regularization scale $\mu$ for the elastic cross section
and leave calculation with hard thermal loop propagators for future studies.

\section*{Acknowledgement}

The authors thank Yu. L. Dokshitzer, C. Greiner and J. Uphoff for
fruitful discussions.  This work was partially supported by the
European Network I3-HP2 Toric, the ANR research program Hadrons@LHC
(grant ANR-08-BLAN-0093-02) and the Pays de la Loire research project
TOGETHER.

\appendix

\section{Spinor QCD matrix element}\label{app:matrix_element}

For completeness we give the matrix elements for spin-$\frac{1}{2}$
quarks,  $\mathcal{M}_1$, $\mathcal{M}_2$ and $\mathcal{M}_3$,
\begin{eqnarray}
\mathcal{M}_1&=& g^3 C_1 \
\frac{\bar{u}(\pthree)\gamma^\mu u(\pb)}{(\pthree-\pb)^2}
D_{\mu\nu}[\pthree-\pb]
\bar{u}(\pone) \frac{\gamma^\nu
(\feyn{P}-\feyn{k}+\mQ)\feyn{\epsilon}}{(\pa-k)^2-\mQsq}
u(\pa)\nonumber\\ 
\mathcal{M}_2&=& g^3 C_2\ \frac{\bar{u}(\pthree)\gamma^\mu
u(\pb)}{(\pthree-\pb)^2} D_{\mu\nu}[\pthree-\pb]
\bar{u}(\pone)\frac{\feyn{\epsilon}
(\feyn{P}'+\feyn{k}+\mQ)\gamma^\nu}{(\pone+k)^2-\mQsq} u(\pa)\nonumber\\
\mathcal{M}_3&=& g^3 C_3
D^{\mu\mu'}[\pthree-\pb]D^{\nu\nu'}[\pone-\pa]
\Big[g_{\mu'\nu'}(\pa-\pone+\pthree-\pb)_\sigma+
g_{\nu'\sigma}(\pone-\pa-k)_{\mu'}+g_{\sigma\mu'}(\pb-\pthree+k)_{\nu'}\Big]
\epsilon^\sigma \nonumber \\ &\times&\frac{\bar{u}(\pthree)\gamma^\mu
u(\pb)\bar{u}(\pone)\gamma^\nu u(\pa)}{(\pthree-\pb)^2(\pone-\pa)^2}.
\end{eqnarray}

\section{Phase space}
\label{app:phase_space}

Here we calculate the three body phase space in the Sudakov variables
of Sect~\ref{sub:diff_x_sec}. We introduce here in addition a finite
gluon mass $m_g$. We assume that the light quark is massless. Using
$p$ and $q$ as defined in Sect~\ref{sub:diff_x_sec} and writing
$k=x\,p+k_q\,q+k_t$, we note first that
\begin{eqnarray}
\int d^3k/(2E_k)&=&\int
d^4k\,\Theta(k^0)\,\delta(k^2-m_g^2)\nonumber\\
&=& \int (p\cdot q)\,dx\,dk_q\,d^2 k_t\,\Theta(x)\,
\delta( 2(p\cdot q)\,x\,k_q - \vkt^2 - m_g^2)\nonumber\\
&=&\int dxd^2 k_t\,\Theta(x)/(2x).
\end{eqnarray}

Introducing $\pthree=x'\,p+\pthree_q\,q+\pthree_t$ and taking into
account $\ell_t=-\pthree_t$, the 3-body phase space reads
\begin{eqnarray} L_3 &=&
\frac{1}{(2\pi)^5}\int d^4\pone\,d^4k\,d^4\pthree
\Theta(\pone^0)\,\delta(\pone^2-\mQsq)
\Theta(k^0)\,\delta(k^2-m_g^2)
\Theta(\pthree^0)\,\delta(\pthree^2)
\delta(\pa+\pb-\pone-k-\pthree)\nonumber \\
&=&
\frac{1}{4\,(2\pi)^5}\int\frac{dx}{x}\Theta(x)d^2 k_t
\frac{dx'}{x'}\Theta(x')d^2 \ell_t
\,\Theta(1-x-x')\delta((\pa+\pb-k-\pthree)^2-\mQsq).
\label{body}
\end{eqnarray}

The integration over $x'$ can be performed after writing
\begin{eqnarray}
\frac{1}{xx'}\delta\bigl((\pa+\pb-k-\pthree)^2-\mQsq\bigr)&=&
\delta\bigl(x'^2(x\,s-\vkt^2-m_g^2)
-x'(x(1-x)\,s-x\,\mQsq-\vkt^2-m_g^2+2x\,\vkt\cdot\vlt)+x(1-x)\vlt^2\bigr)
\nonumber\\
&=&\frac{\Theta(\Delta)}{\sqrt{\Delta}}
\bigl\{\delta(x'-x'_1)+\delta(x'-x'_2)\bigr\},
\end{eqnarray}
with
\begin{equation}
\Delta=\bigl(x(1-x)\,s-x\,\mQsq-\vkt^2-m_g^2+2x\,\vkt\cdot\vlt\bigr)^2
-4x(1-x)\,\vlt^2\,(x\,s-\vkt^2-m_g^2).
\end{equation}
Among the two roots, $x'_1$ and  $x'_2$, one is of order 1 and the
other is of order $\vlt^2/s$. Only the latter is relevant at
high-energy, because of the suppression by the elastic matrix element
at large $|t|$ and since $t=\ell^2=(q-q')^2=-x'\,(2p\cdot q)$. For
completeness we also give the boundaries in explicit forms:
\begin{eqnarray}
0\le x &\le& 1-\mQsq/s,\nonumber\\
\vkt^2 &\le& x(1-x)\,s-x\,\mQsq-m_g^2,\nonumber\\
\label{eq:boundaries}
|\vlt| &\le&\frac{x(1-x)\,s-x\,\mQsq-\vkt^2-m_g^2}%
{2\left(\sqrt{x(1-x)(x\,s-\vkt^2-m_g^2)}-x\,|\vkt|\cos\phi\right)}.
\end{eqnarray}

\section{Comparison between the QCD and the QED-like gluon
  distributions}
\label{app:qed_vs_qcd}

Here we want to further investigate the comparison between the gluon
distribution eq.~\ref{eq:gluon_distribution} as deduced from
eq.~\ref{eq:M2SQCD} and the QED-like one that could be deduced
from eq.~\ref{eq:M2SQED}. Our aim is to extract the gross features in
limiting regimes in order to prove that the QED-like terms contribute
only little to radiation. The discussion thus follows from the one
given when commenting Fig.~\ref{fig:gluon_kt}.

Again the discussion is carried out for $\mathcal{I}$, the
$\phi_{k_t}$-integrated gluon distribution, which is made explicit in
eq.~\ref{gbia} and will be referred to as the QCD distribution. 
The QED-like distribution is deduced from eq.~\ref{gbia}
by changing $\vlt\to x\,\vlt$ and $C_A\to C_F$. For simplicity the
distributions will be shown omitting their color factors and the
common prefactor $\frac{2 \alphas}{\pi}\frac{1-x}{x}$. Since we are
mostly interested in the comparison between the QCD and the QED-like
distributions the common prefactor is inessential.

The parameter space $(\lt,\mQ)$ can be split into two parts. When
$\mQ<\lt$ the regime is hard for every $x$ for the QCD
distribution (since $\mQ<\lt\Rightarrow x\,\mQ<\lt$). We recall that
in such regime we have (assuming a strong hierarchy of scales)
\begin{itemize}
\item $\mathcal I\propto 1/\vlt^2$ (here and in the following the
  constant of proportionality is the color factor times the common
  prefactor which we decided to skip) at small $\kt$, $\kt\ll
  (x\mQ)^2/\lt$;
\item $\mathcal I\propto \vkt^2/(x\mQ)^4$ in the dead-cone region
  $(x\mQ)^2/\lt\ll\kt\ll x\mQ$;
\item $\mathcal I\propto 1/\vkt^2$ in the log-enhanced region
  $x\mQ\ll\kt\ll\lt$;
\item $\mathcal I\propto \vlt^2/(\vkt^2)^2$ at large $\kt$,
  $\kt\gg\lt$.
\end{itemize}
\begin{figure}[H]
\begin{center}
\epsfig{file=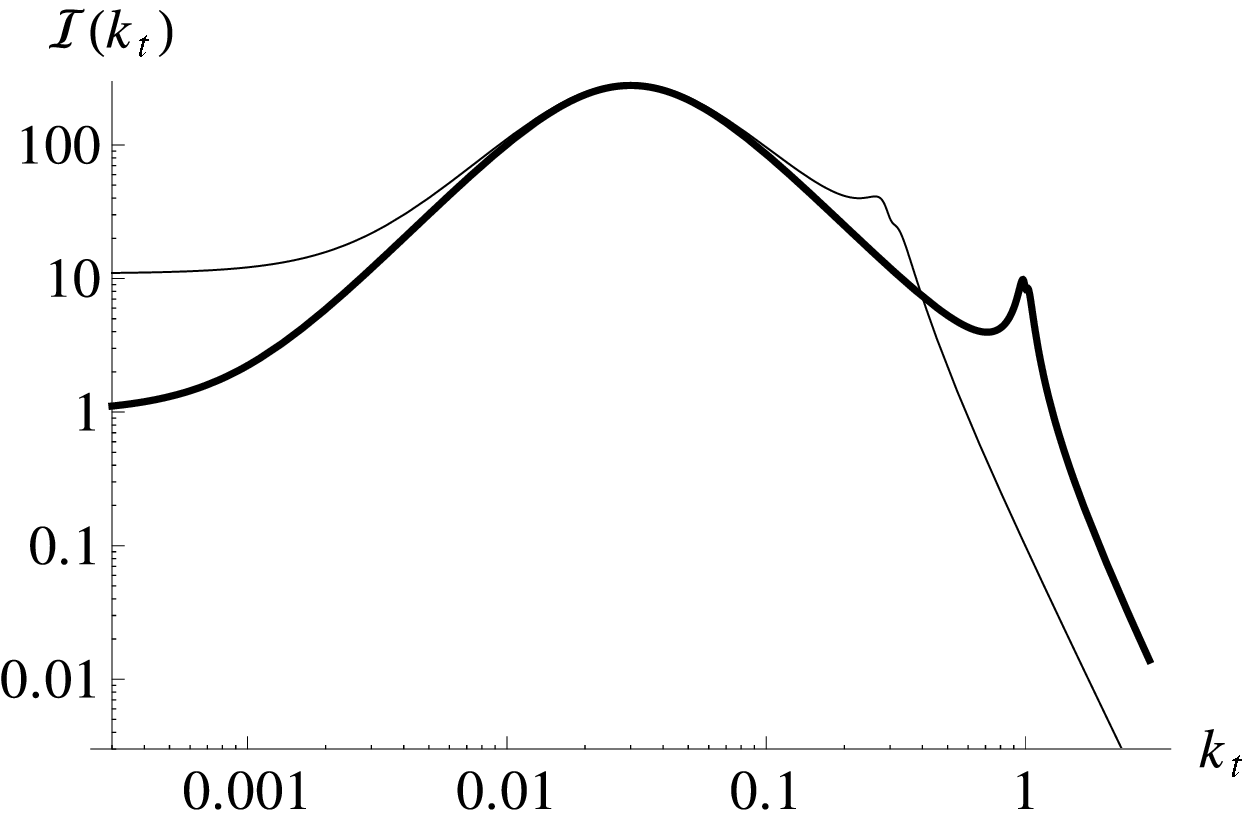,width=0.45\textwidth}
\epsfig{file=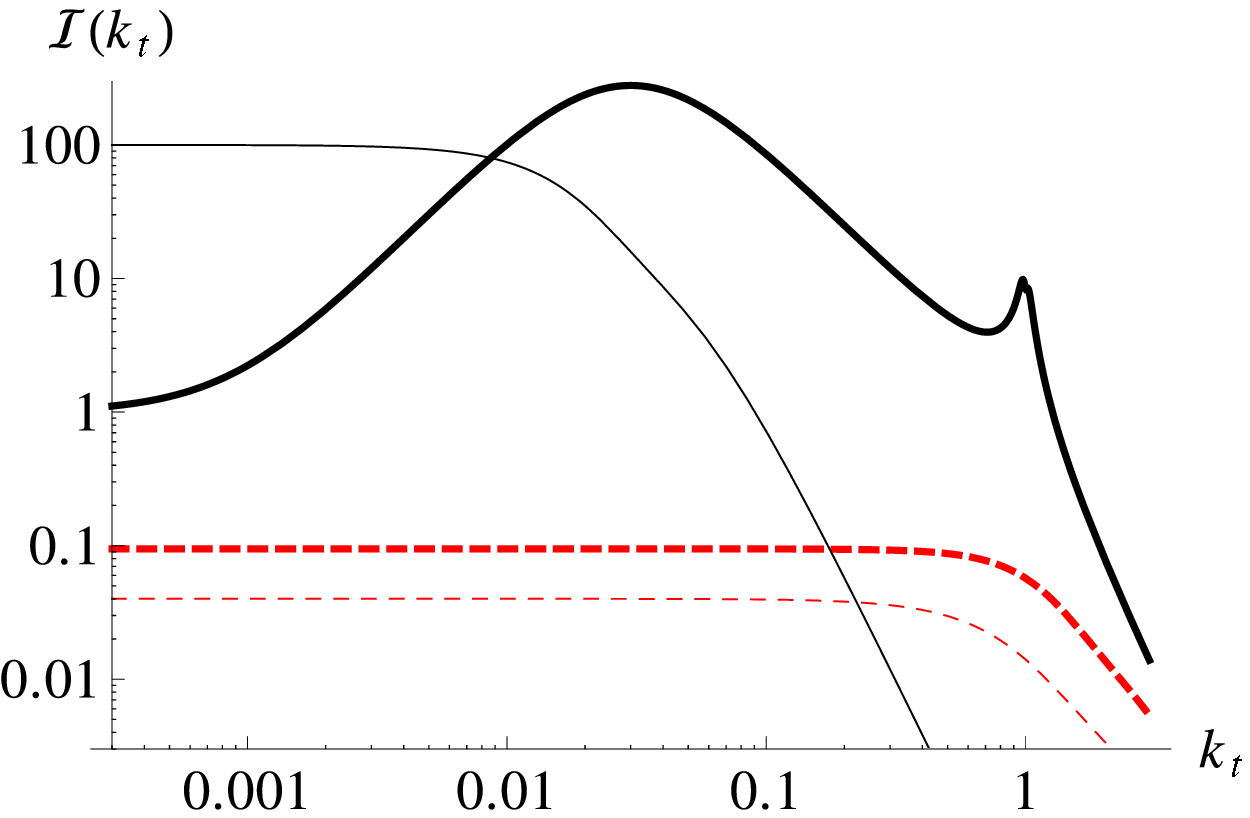,width=0.45\textwidth}
\end{center}
\caption{(Color online) $\kt$ distribution of gluons
  $\mathcal{I}(x,|\vkt|,t;\mQ)$ for the QCD (thick lines) and the
  QED-like cases. Left (hard scattering only): $|\vlt|=1$, $\mQ=0.1$ and $x=0.3$. Right :
  $|\vlt|=1$, $\mQ=3.$; $x_1=0.5$ (dashed lines; soft scattering) and $x_1=0.01$ (solid lines).}
\label{fig:gluon_c_and_e}
\end{figure}

This trend is visible in Fig.~\ref{fig:gluon_c_and_e} (left) which
shows the QCD distribution (thick line). The QED-like
distribution is also shown for comparison (thin line). For $\mQ<\lt$
the regime is also hard for the latter, since
$x\,\mQ<x\,\lt$. Therefore the same sequence of behavior is
obtained but the change of $\lt$ by $x\,\lt$ squeezes the intermediate
range. Both curves are on top of each other in both, the dead-cone and
the log-enhanced windows, since the function there is
$\lt$-independent. With the above-mentioned squeezing the log-enhanced
window is shorten in the QED-like case and therefore the QED-like
distribution becomes negligible for $\kt>x\,\lt$. A further
consequence is that the $\kt$-integrated distribution is
$\propto\ln(\lt/M)$ for the QED-like case to be compared with
$\ln(\lt/(x\,M))$ for the QCD case, hence the dominance of the
QCD distribution at small $x$. At small $\kt$ the squeezing of
the dead-cone window results in the dominance of the QED distribution
which is $\propto 1/(x^2\vlt^2)$. This region is irrelevant for the
whole radiation as the $\kt$-integration demonstrates.

When $\mQ>\lt$, the regime is soft for the QED-like distribution, as
$x\,\lt<x\,\mQ$. For the QCD distribution, as thoroughly
investigated in the main part of the paper, the regime is hard at
small $x$ when $x\,\mQ<\lt$ and becomes soft for larger $x$. We recall
that the soft regime for the QCD distribution is characterized by
\begin{itemize}
\item $\mathcal I\propto \vlt^2/(x\,\mQ)^4$ at small $\kt$, $\kt\ll
  x\,\mQ$;
\item $\mathcal I\propto \vlt^2/(\vkt^2)^2$ at large $\kt$,
  $\kt\gg x\,\mQ$.
\end{itemize}
The various possibilities are plotted in Fig.~\ref{fig:gluon_c_and_e}
(right). The thick red dashed line is the QCD distribution for a typical
large $x$ value, $x=x_1$ (soft regime), and the thin red dashed line is 
the QED-like distribution for the same $x=x_1$. From the above behaviors
and the change $\lt\to x\,\lt$ when going from QCD to QED-like we
see that the QED-like distribution is simply scaled down by a factor
$x^2$ with respect to the QCD one (property C). 
This simple scaling translates
to the $\kt$-integrated distributions, being respectively
$\propto\vlt^2/(x\,\mQ)^2$ and $\vlt^2/\mQsq$ for the QCD and QED-like
situations. At smaller $x$, $x=x_2$, the trend is exemplified in
Fig.~\ref{fig:gluon_c_and_e} (right) with the thick black curve
(QCD) and the thin black curve (QED-like). For $x<\vlt^2/\mQsq$
($x_2$ was chosen in this range), the QED-like distribution overshoots
the QCD one at small $\kt$, precisely when $\kt<x\,\lt$. This is
the property B pointed out in Sect.~\ref{sub:small_kt} -- with $l_t$ taken 
as $\mD$ -- when discussing the ratio ${\bar P}_g^{SQED}/{\bar P}_g^{tot}$. 
As already mentioned this overshooting is harmless for the overall radiation. For
$\vlt^2/\mQsq<x<\lt/\mQ$, the curves (not plotted) show the same trend
as the solid curves in Fig.~\ref{fig:gluon_c_and_e} (right) but the
QCD distribution is always greater than the QED-like one, with 
however a sizable ratio $\mathcal{O}\left(\frac{l_t^2}{x M^2}\right)^2$
provided $k_t\lesssim \frac{(x M)^2}{l_t}$ (property A).


\begin{thebibliography}{11}
\bibitem{Borsanyi:2010cj}
  S.~Borsanyi, G.~Endrodi, Z.~Fodor, A.~Jakovac, S.~D.~Katz, S.~Krieg, C.~Ratti and K.~K.~Szabo,
  JHEP {\bf 1011}, 077 (2010).

\bibitem{Andronic:2007}
A. Andronic, P. Braun-Munzinger, K. Redlich, J. Stachel,
Nucl.Phys.A {\bf 789}, 334 (2007).


\bibitem{Vogt:2003}
R. Vogt, Int. J. Mod. Phys. E {\bf 12}, 211 (2003). 

\bibitem{Cacciari:2005}
M. Cacciari, P. Nason, R. Vogt, Phys. Rev. Lett. {\bf 95}, 122001 (2005)  

\bibitem{Cacciari:2012}
M. Cacciari, S.~Frixione, N.~Houdeau, M.~L. Mangano, P.~Nason, G.~Ridolfi,
JHEP {\bf 10}, 137 (2012).

\bibitem{Phenix:2011}
PHENIX Collaboration (A. Adare et al.),
Phys. Rev.C {\bf 84}, 044905 (2011).

\bibitem{ALICE:2012}
ALICE Collaboration (B. Abelev  et al.),
JHEP {\bf 09}, 112 (2012). 

\bibitem{Gossiaux:2011ea}
  P.~B.~Gossiaux, S.~Vogel, H.~van Hees, J.~Aichelin, R.~Rapp, M.~He
  and M.~Bluhm, arXiv:1102.1114 [hep-ph].
  
\bibitem{Bjorken} J.~D.~Bjorken, Fermilab preprint Pub-82/59-THY (1982). 

\bibitem{Braaten:1991jj}
E.~Braaten and M.~H.~Thoma, Phys.\ Rev.\  D {\bf 44}, 1298 (1991);
Phys.\ Rev.\ D {\bf 44}, 2625 (1991). 

\bibitem{Peshier:2006hi}
  A.~Peshier,
  Phys.\ Rev.\ Lett.\  {\bf 97}, 212301 (2006).

\bibitem{Peigne:2008nd}
  S.~Peigne and A.~Peshier,
  Phys.\ Rev.\ D {\bf 77}, 114017 (2008). 

\bibitem{Gossiaux:2008jv}
  P.~B.~Gossiaux and J.~Aichelin,
  Phys.\ Rev.\  C {\bf 78}, 014904 (2008);
  J.\ Phys.\ G {\bf 36}, 064028 (2009).

\bibitem{Gyulassy94} 
M.~Gyulassy and X.-N.~Wang, Nucl. Phys. B {\bf 420}, 583 (1994).

\bibitem{Wang95} 
X.-N.~Wang, M.~Gyulassy, and M.~Pl\"umer, Phys. Rev. D {\bf 51}, 3436 (1995).

\bibitem{Baier95} R.~Baier, Y.~L.~Dokshitzer, S.~Peign\'{e}, and D.~Schiff, 
 Phys. Lett. B {\bf 345}, 277 (1995). 

\bibitem{Baier97} 
R.~Baier, Y.~L.~Dokshitzer, A.~H.~M\"uller, S.~Peign\'{e}, and D.~Schiff, 
  Nucl.\ Phys.\ B {\bf 478}, 577 (1996); ibid. {\bf 483}, 291 (1997); ibid. {\bf 484}, 265 (1997). 


\bibitem{Zakharov} 
B.~G.~Zakharov, JETP Lett. {\bf 63},952 (1996) ;
 ibid. {\bf 65}, 615 (1997); ibid. {\bf 73}, 49 (2001);
 Phys.\ Atom.\ Nucl.\  {\bf 61}, 838 (1998) [Yad.\ Fiz.\  {\bf 61}, 924 (1998)].

\bibitem{GLV} 
M.~Gyulassy, P.~Levai, and I.~Vitev, Phys. Rev. Lett. {\bf 85}, 5535 (2000); 
Nucl. Phys. B {\bf 571}, 197 (2000); ibid. {\bf 594}, 371 (2001). 

\bibitem{Dokshitzer:2001zm} 
Y.~L.~Dokshitzer and D.~E.~Kharzeev, Phys. Lett. B {\bf 519}, 199 (2001). 

\bibitem{AMY} 
P.~B.~Arnold, G.~D.~Moore, and L.~G.~Yaffe, JHEP {\bf 0011}, 001 (2000); 
 ibid. {\bf 0305}, 051 (2003). 
 
 
\bibitem{Zakharov:2004}
B. G. Zakharov, JETP Lett. {\bf 80}, 617 (2004)  
 
\bibitem{ASW} 
N.~Armesto, C.~A.~Salgado, and U.~A.~Wiedemann, Phys. Rev. D {\bf 69}, 114003 (2004); 
Phys. Rev. C {\bf 72}, 064910 (2005). 

    
\bibitem{Zhang:2004}
B.-W. Zhang, E.Wang and X.-N.Wang, 
 Phys. Rev. Lett. {\bf 93}, 072301 (2004).

\bibitem{Djordjevic:2004}
M. Djordjevic and M. Gyulassy, Nucl.Phys. A {\bf 733}, 265 (2004); 
M. Djordjevic, M. Gyulassy and S.Wicks, Phys. Rev. Lett. {\bf 94}, 112301 (2005). 


\bibitem{Wicks:2007am}
  S.~Wicks, W.~Horowitz, M.~Djordjevic and M.~Gyulassy,
  Nucl.\ Phys.\ A {\bf 783}, 493 (2007) 



\bibitem{Baier:2000mf}
  R.~Baier, D.~Schiff and B.~G.~Zakharov,
  Ann.\ Rev.\ Nucl.\ Part.\ Sci.\  {\bf 50}, 37 (2000).

\bibitem{Baier:1998yf}
  R.~Baier, Y.~L.~Dokshitzer, A.~H.~Mueller and D.~Schiff,
  Phys.\ Rev.\  C {\bf 58}, 1706 (1998).
    
\bibitem{Zakharov:98}
B.G. Zakharov,  arxiv :hep-ph/9807396v1.
    
\bibitem{Arnold:2002ja}
  P.~B.~Arnold, G.~D.~Moore and L.~G.~Yaffe,
  JHEP {\bf 0206}, 030 (2002).


\bibitem{Gossiaux:2009mk}
  P.~B.~Gossiaux, R.~Bierkandt and J.~Aichelin,
  Phys.\ Rev.\  C {\bf 79}, 044906 (2009).

\bibitem{Kolb:2003dz}
  P.~F.~Kolb and U.~W.~Heinz,
  In *Hwa, R.C. (ed.) et al.: Quark gluon plasma*, 634-714

\bibitem{Gossiaux:2010yx}
  P.~B.~Gossiaux, J.~Aichelin, T.~Gousset and V.~Guiho,
  Phys. G {\bf 37}, 094019 (2010).
  

\bibitem{Gunion:1981qs}
  J.~F.~Gunion and G.~Bertsch, Phys.\ Rev.\  D {\bf 25}, 746 (1982).
  

\bibitem{Fochler:2013}
O.~Fochler, J.~Uphoff, Z.~Xu and C.~Greiner, arXiv:1302.5250.


\bibitem{Djordjevic:2007at}
  M.~Djordjevic and U.~Heinz,
  Phys.\ Rev.\ C {\bf 77}, 024905 (2008).


\bibitem{Werner:2010aa}
  K.~Werner, I.~Karpenko, T.~Pierog, M.~Bleicher and K.~Mikhailov,
  Phys.\ Rev.\ C {\bf 82}, 044904 (2010).

\bibitem{Werner:2012xh}
  K.~Werner, I.~Karpenko, M.~Bleicher, T.~Pierog and S.~Porteboeuf-Houssais,
  Phys.\ Rev.\ C {\bf 85}, 064907 (2012). 

\bibitem{gosssiaux:qnp}
P. B. Gossiaux, J. Aichelin, M. Bluhm, T. Gousset,
M. Nahrgang, S. Vogel and K. Werner, PoS QNP 2012, 160 (2012).


\bibitem{IZ}
  C. Itzykson and J.-B. Zuber, Quantum Field Theory, Dover
  Publications, Mineola, New York (2005).

\bibitem{Meggiolaro:1995cu}
  E.~Meggiolaro, Phys.\ Rev.\ D {\bf 53}, 3835 (1996).

\bibitem{Silva:2013}
P. J. Silva, O. Oliveira, P. Bicudo, N. Cardoso, arXiv:1310.5629.

\bibitem{Peshier:2007}
A. Peshier, arxiv: hep-ph/0601119.

\bibitem{Peigne:2008wu}
  S.~Peigne and A.~V.~Smilga,
  Phys.\ Usp.\  {\bf 52}, 659 (2009)
  [Usp.\ Fiz.\ Nauk {\bf 179}, 697 (2009)].
  
  \bibitem{Baier:2001yt}
    R.~Baier, Y.~L.~Dokshitzer, A.~H.~Mueller and D.~Schiff,
    JHEP {\bf 0109}, 033 (2001).  

\end{thebibliography}
\end{document}